\documentclass[11pt,a4paper]{article}
\pdfoutput=1

\usepackage[a4paper,text={16.8cm,22.4cm}]{geometry}
\usepackage{amsmath,amsfonts,slashed,amssymb,tikz,bm,psfrag,graphicx,color,dsfont}
\usepackage{multicol}
\usepackage{float}
\usepackage{slashed}
\usepackage{euscript}
\usepackage[normalem]{ulem}
\RequirePackage[sort&compress,square,comma,numbers]{natbib}
\allowdisplaybreaks
\renewcommand{\arraystretch}{1.2}
\usepackage[utf8]{inputenc}
\usepackage[T1]{fontenc}
\usepackage{graphicx}
\usepackage{amsmath}
\usepackage{amsfonts}      
\usepackage{empheq}       
\usepackage{hyperref}     
\usepackage{euscript}
\usepackage{multirow}
\usepackage{simplewick}
\usepackage{xcolor}
\usepackage{slashed}
\usepackage{rotating} 
\usepackage{pgfplots}
\pgfplotsset{compat=1.18}
\usepackage{filecontents}
\usepackage[utf8]{inputenc}
\usepackage{placeins}

\DeclareUnicodeCharacter{200B}{}

\allowdisplaybreaks

\definecolor{brown}{RGB}{150,50,0}

\begin{document}

\begin{titlepage}

\vspace{0.1cm}
\begin{center}
\Large\bf
Exclusive Determination of $|V_{cb}|$ from Semileptonic Decays $B\to D^{(*)}\ell \nu_{\ell}$
\end{center}

\vspace{0.5cm}
\begin{center}

{\bf Xue-Wen Chen$^a$\,, Bo-Yan Cui$^a$\,, Jia-Wei Zhang$^a$\,, Xue-Chen Zhao$^{\ast,b}$\,, Ya-Hui Chen$^{\dagger,c}$} \\
\vspace{0.5cm}
{\sl
${}^a$ Department of Physics, Chongqing University of Science and Technology,\\
Chongqing 401331, P.R. China
\\
${}^b$ School of Physics, Nankai University, 300071 Tianjin, P.R. China \\
${}^c$Department of Physics, College of Basic Medical Sciences, Army Medical University,
Chongqing 400038, P.R. China
}
\\

\vspace{0.2cm}
{\sl E}-{\it mail}: {\tt 
chenxuewen@cqust.edu.cn,\quad boyancui@cqust.edu.cn,\quad jwzhang@cqust.edu.cn, \\
zxc@mail.nankai.edu.cn,\quad cyh@tmmu.edu.cn}
\vspace{0.5cm}

\end{center}


\begin{abstract}

We present an updated exclusive determination of the CKM matrix element \(|V_{cb}|\) from the semileptonic decays \(B\to D^{(*)}\ell\bar{\nu}_{\ell}\). 
Our analysis combines the latest Belle II measurements, updated lattice-QCD calculations of the \(B\to D^{(*)}\) form factors at small hadronic recoil, 
and correlated large-recoil SCET sum-rule predictions incorporating next-to-leading-order QCD corrections and several power-suppressed contributions. We 
consider three fit scenarios with progressively enlarged input sets and find that the inclusion of the large-recoil sum-rule constraints substantially 
reduces the form-factor uncertainties. From the full global fit, we obtain \(|V_{cb}|=(39.18\pm0.47)\times10^{-3}\). Using the combined lattice-QCD and 
LCSR fit, we predict \(R(D)=0.3069\pm0.0080,\qquad R(D^*)=0.2548\pm0.0043\), 
and provide differential decay distributions in the momentum transfer and angular variables for both the muon and tau channels. Comparisons of the 
individual and correlated predictions for \(R(D)\) and \(R(D^*)\) with the experimental averages reveal a persistent tension. In particular, our 
theoretical 68\% confidence region shows little overlap with the experimental average. All correlations among the fitted parameters are retained in the 
uncertainty propagation. Our results therefore provide updated Standard Model benchmarks for tests of lepton-flavour-universality. Improved lattice-QCD 
calculations, sum-rule predictions, and Belle II measurements will be essential for determining whether the remaining discrepancies originate from 
theoretical systematic uncertainties or from physics beyond the Standard Model.
\end{abstract}
\vfil

\begin{flushleft}
\normalsize

$^\ast$Corresponding author.\\
$^{\dagger}$Corresponding author.
\end{flushleft}
\end{titlepage}

\tableofcontents

\setcounter{tocdepth}{2}

\makeatletter
\gdef\@fpheader{}
\makeatother
\flushbottom

\section{Introduction}\label{sec:introduction}

The Cabibbo-Kobayashi-Maskawa (CKM) matrix~\cite{Cabibbo:1963yz,Kobayashi:1973fv} describes the flavour mixing of quarks  in the Standard Model. 
The CKM matrix element $|V_{cb}|$ quantifies the transition strength between the charm and bottom quark in charged weak interactions. 
It is used to verify the unitarity of the CKM matrix and to search for new physics beyond the Standard Model (SM). 
A precise determination of $|V_{cb}|$ is one of the most important goals of B-physics phenomenology.
The discrepancy between exclusive and inclusive extractions of $|V_{cb}|$  has motivated extensive studies to solve this puzzle in 
the Standard Model and beyond~\cite{Belle-II:2018jsg,Gambino:2020jvv,Boyle:2022uba,USQCD:2022mmc}. 

The Belle collaboration measured the semileptonic decay $B \to D \ell \nu_{\ell}$ via fully reconstructed $B$ meson 
tag events~\cite{Glattauer:2015teq}, obtained its isospin-averaged branching fraction, and extracted the CKM matrix element $|V_{cb}|$ 
with two parameterization frameworks. Binned differential decay rates in bins of the recoil variable $w$ were also presented. 
The Belle II Collaboration performed an analogous measurement in Ref.~\cite{Belle-II:2025rna}, adopting the BCL parameterization. 
The determination of $|V_{cb}|$ can also be performed via $B^{0}\to D^{*-}\ell^{+}\nu_{\ell}$ by carrying out independent fits using 
the CLN and BGL form factor parameterizations\cite{Belle:2018ezy}, alongside full binned differential distributions, response matrices, 
and statistical as well as systematic correlation matrices to alleviate the long-standing inclusive–exclusive $|V_{cb}|$ tension. 
In Ref.~\cite{Belle-II:2023okj}, Belle II Collaboration perform the determination of $|V_{cb}|$ by using 
$\bar{B}^{0}\to D^{*+}\ell^{-}\bar{\nu}_{\ell}$ decays, the separate partial decay rates for electron and muon final states 
are also provided as functions of the recoil parameter $\omega$ and three decay angular observables.  These experimental analyses require 
lattice-QCD form factors evaluated near zero recoil as external inputs to the fits for extracting 
the CKM matrix element $|V_{cb}|$. Lattice QCD simulations of $B\to D^{(*)}\ell\bar{\nu}$ semileptonic form factors, 
the core input to extract the CKM matrix element $|V_{cb}|$, have undergone steady methodological and precision improvements over the past decade. 
The first unquenched lattice gauge simulation for the $B \to D\ell \bar{\nu}$ hadronic form factors was presented by Fermilab-MILC 
collaborations~\cite{FermilabLattice:2015ilb}. Then, the HPQCD Collaboration performs full-$q^2$ unquenched lattice QCD calculations of 
the $B\to D\ell\nu$ semileptonic form factors $f_+(q^2)$ and $f_0(q^2)$~\cite{Na:2015kha}. For the exclusive mode $B\to D^* \ell \nu$, 
the unquenched lattice-QCD calculation of the form factors at nonzero recoil was performed by Fermilab-MILC collaborations with $N_f=2+1$ flavours~\cite{FermilabLattice:2021cdg}. HPQCD produced complete vector, axial-vector, and tensor form factors for $B\to D^*\ell\nu$ and 
$B_s\to D_s^*\ell\nu$ covering the entire $q^2$ range, enriching theoretical inputs for rare and standard semileptonic 
observables~\cite{Harrison:2023dzh}. Meanwhile, JLQCD introduced M\"obius domain-wall fermion discretization for $B\to D^*\ell\nu_\ell$ 
transitions~\cite{Aoki:2023qpa}, offering an independent formulation cross-check against the widely used clover heavy-quark schemes of MILC 
and HPQCD. Collectively, these works trace a trajectory from narrow zero-recoil single-channel studies toward full-$q^2$, multi-flavour, 
multi-operator form factor sets with multiple independent lattice discretization frameworks, drastically reducing lattice 
systematic uncertainties in $|V_{cb}|$ extraction. In view of the fact that including large hadronic recoil light-cone sum rule (LCSR) predictions 
in numerical fits can substantially reduce fitting uncertainties~\cite{Wang:2017jow,Gao:2021sav,Cui:2022zwm,Cui:2023jiw,Gao:2024vql,Khodjamirian:2023wol}, 
the high precision calculation especially the next-to-leading-order (NLO) and next-to-leading-power (NLP) for exclusive semileptonic 
$\bar{B}\to D^{(*)}\ell \bar{\nu}_{\ell}$ form factors will therefore be in high demand for pinning down the obtained uncertainties of their 
shape parameters from the $z$-series fitting procedure. Combined numerical fit incorporating the LCSR predictions at large hadronic recoil, 
the lattice QCD simulations at zero/nonzero recoil and the available experimental data 
points~\cite{Bigi:2017jbd,Bordone:2019guc,Jaiswal:2017rve,Jaiswal:2020wer,Biswas:2022yvh,Cheung:2020sbq,Fang:2026hru} was performed. 
Inspired by the latest lattice QCD calculations from HPQCD~\cite{Harrison:2023dzh} and JLQCD~\cite{Aoki:2023qpa}, alongside high-precision 
Belle II measurements of $B\to D\ell\nu$~\cite{Belle-II:2025rna} and $B\to D^*\ell\nu$~\cite{Belle-II:2023okj}, a joint fit is essential to 
yield the most precise determination of the CKM matrix element $|V_{cb}|$ available to date. In this work, we will perform 
a joint fit incorporating all the aforementioned state-of-the-art inputs to extract the CKM matrix element $|V_{cb}|$, 
and derive Standard Model predictions for several observables of experimental interest. 

The remainder of this paper is organized as follows: we will give the input data point, the BGL parameterization and strong unitarity bound at 
section~\ref{sec:framework}, the fitted result and some Standard Model predictions for several observables are shown 
in section~\ref{sec:result}, finally, a brief conclusion is given at section~\ref{sec:conclusion}.

\section{Fitting framework}\label{sec:framework}
\subsection{$B\to D^{(*)}$ form factors}
The $B\to D^{(*)}$ form factors are defined by
\begin{eqnarray}
\langle D(p)|\bar{c} \, \gamma_{\mu} \, b |\bar{B}(p+q) \rangle
&=& f_+(q^2)\left[2p_{\mu}+\left(1-\frac{m_B^2-m_D^2}{q^2} \right)q_{\mu} \right] \nonumber \\
&+& f_0(q^2)\left(\frac{m_B^2-m_D^2}{q^2} \right)q_{\mu}\,, 
\\
   \langle D^*(p, \epsilon^{\ast}) | \bar{c} \, \gamma_{\mu} \, b | \bar B (p+q) \rangle
&=& - \frac{2 \, i \,   V(q^2) }{ m_{B} + m_{D^*}} \,
\epsilon_{\mu \nu \rho \sigma} \, \epsilon^{\ast \, \nu}\,
p^{\rho} \, q^{\sigma} \,, 
\\
 \langle D^*(p, \epsilon^{\ast}) | \bar{c} \, \gamma_{\mu} \, \gamma_5 \, b | \bar B (p+q)\rangle
&=&  \frac{2 \, m_{D^*} \, \epsilon^{\ast} \cdot q  }{ q^2} \, q_{\mu} \, A_0(q^2) \nonumber \\
&+& (m_{B} +m_{D^*})\, \left[\epsilon^{\ast}_{\mu} - \frac{\epsilon^{\ast} \cdot q }{ q^2}  \, q_{\mu} \right ]
 \, A_1(q^2)    \nonumber  \\
&-& \frac{\epsilon^{\ast} \cdot q  }{ m_{B} +m_{D^*}} \, \left [ (2 \, p + q)_{\mu}
- {m_{B}^2 - m_{D^*}^2 \over q^2}  \, q_{\mu}\right ] \, A_2(q^2) \,, 
\end{eqnarray}
with the convention $\epsilon_{0123}=-1$. At maximal hadronic recoil $q^2=0$, the following relations among the form factors hold:
\begin{eqnarray}
f_{+}^{BD}(0) &=& f_{0}^{BD}(0) \label{fpeqf0} \,,\\
A_0(0)&=&{m_{B} + m_{D^*} \over 2 \, m_{D^*}} \, A_1(0) - {m_{B} - m_{D^*} \over 2 \, m_{D^*}} \, A_2(0)  \,,\label{A0eqA1-A2}
\end{eqnarray}
which are free of radiative and power corrections. 

\subsection{BGL Parametrization}\label{sec:BGL}

The BGL parametrization introduces a function which is dependent on the momentum transfer squared $q^{2}$ \cite{Boyd:1997kz}

\begin{eqnarray}
	z(q^{2},t_{0})=\frac{\sqrt{t_{+}-q^{2}}-\sqrt{t_{+}-t_{0}}}{\sqrt{t_{+}-q^{2}}+\sqrt{t_{+}-t_{0}}} \, , \qquad 
t_{\pm}=\left(M \pm m\right)^{2} \, ,
\end{eqnarray}
where $M,m$ refer to the masses of beauty hadron and charm hadron respectively, and $t_{0}$ is a free parameter. $t_{+}$ is the pair-production 
threshold, and $t_{-}$ is the physical kinematic upper bound of $q^{2}$ in the $b\to c$ semileptonic decays. For a heavy-to-heavy transitions, 
it is convenient to introduce a variable $\omega$
\begin{eqnarray}
	\omega\equiv v\cdot v'=\frac{M^{2}+m^{2}-q^{2}}{2 M m} \, ,
\end{eqnarray}
where $v=p/M$ and $v'=p'/m$ are velocities of initial state (with momentum $p$) and final state (with momentum $p'$). In the $b$ rest frame, 
$\omega$ depends on the energy transfer to the light degrees of freedom, and relates $b\to c$ form factors in the heavy quark limit. 
The variable $z(q^{2},t_{0})$ shows the same property obviously by this form
\begin{eqnarray}
	z(\omega,N)=\frac{\sqrt{1+\omega}-\sqrt{2N}}{\sqrt{1+\omega}+\sqrt{2N}} \, , \quad 	N=\frac{t_{+}-t_{0}}{t_{+}-t_{-}} \, .
\end{eqnarray}
$N$ is a free parameter depends on $t_{0}$, and we set $N=1$ as default in the later content, which corresponding to $t_{0} = t_{-}$. 
The $B\to D^{(*)}$ four processes depend on the same argument $z(\omega,N)$, but the meson masses $M$ and $m$ should be consistent with 
the special process. Introducing the Blaschke factors $P_{i}$ as well as the outer functions $\phi_{i}$, form factors are parametrized as
\begin{eqnarray}
	f_{i}(z)&=&\frac{1}{P_{i}(z) \, \phi_{i}(z)}\sum^{\infty}_{j=0} b_{j}^{i} z^{j} \, ,
	\label{BGLpara}
\end{eqnarray}
where $i$ refers to different form factor, and $f_{i} \in \{f_{+},f_{0},g,f,{\cal F}_{1},{\cal F}_{2}\}$. $g,f,{\cal F}_{1},{\cal F}_{2}$ are form factors of $B\to D^{*}$ defined by BGL parametrization, which can be obtained by the transformations
\begin{eqnarray}
	\begin{split}
		g&= \frac{2}{m_{B}(1+r_*)} V \, , \quad f= m_{B}(1+r_*) A_{1} \, , \quad \mathcal{F}_{2}= 2 A_0 \, ,\\
		\mathcal{F}_{1}&= m_{B}^2 (1+\omega)\left[\frac{1+r_*}{1+\omega}\left(\omega-r_*\right) A_{1}
		+\frac{2r_*}{1+r_*}(1-\omega)A_{2}\right] \, .
	\end{split}
	\label{BGLFFs}
\end{eqnarray}
where $r_*=m_{D^*}/m_B$. Substituting equation~\eqref{A0eqA1-A2} into \eqref{BGLFFs}, one can easily derive at $q^{2}=0$ or 
$\omega=\omega_{\text{Max}}=(1+r_*^2)/(2r_*)$,
\begin{eqnarray}
	\mathcal{F}_{2}(\omega_{\text{Max}})=\frac{1+r_*}{m^{2}_{B}(1+\omega_{\text{Max}})(1-r_*)r_*}\mathcal{F}_{1}(\omega_{\text{Max}}) \, ,
	\label{F2=F1}
\end{eqnarray}
and implied by \eqref{BGLFFs}
\begin{eqnarray}
	\mathcal{F}_{1}(\omega=1)=m_{B}(1-r_*)f(\omega=1) \, .
	\label{F1=f}
\end{eqnarray}

The Blaschke factors $P_{i}$ remove poles produced by $B_{c}$ resonances below the pair-production threshold, in the $q^{2}$-space $ P_{i}(q^{2}) = \prod_{p} z(q^{2},m_{p}^{2})$, and in the $z$-space

\begin{eqnarray}
	P_{i}(z)=\prod_{p}\frac{z-z_{p}}{1-z \, z_{p}}\, , \quad z_{p}\equiv z(m_{p}^{2},t_{0}) \, ,
\end{eqnarray}
where the index $p$ refer to the resonances whose quantum numbers are the same as form factor $f_{i}$. Relevant $B_{c}$ resonance masses $m_{p}$ 
are collected in Table~\ref{Bc_resonances}. Vector form factors $\{f_{+},g\}$ correspond with $1^{-}$ states, $\{f,{\cal F}_1\}$ with $1^{+}$, 
${\cal F}_2$ with $0^{-}$, and $f_{0}$ with $0^{+}$. $B_{c}(3 \, ^{3} \! S_{1}) (7.280)$ is very close to the threshold for $B \to D^{*}$, 
in Ref.\cite{Cui:2023jiw} this pole is included in parametrization, while in this work we also consider this resonance in $B \to D^{*}$ 
process for the convenience of comparing our results with other recent work. 

\begin{table}[htbp]
	\centering
	\renewcommand{\arraystretch}{1.5}
	\begin{tabular}{ccc|ccc}
		\hline\hline
		Type & Mass (GeV) & ref. & Type & Mass (GeV) & ref. \\
		\hline
		$1^{-}$ & $6.329(3)$ & \cite{ParticleDataGroup:2016lqr,Dowdall:2012ab,Colquhoun:2015oha} & $1^{+}$ & $6.793(13)$ & \cite{Dowdall:2012ab} \\
		$1^{-}$ & $6.920(18)$ & \cite{Dowdall:2012ab,Ikhdair:2005xe} & $1^{+}$ & $6.750$ & \cite{Godfrey:2004ya} \\
		$1^{-}$ & $7.020$ & \cite{Rai:2013xvr} & $1^{+}$ & $7.145$ & \cite{Godfrey:2004ya} \\
		$1^{-}$ & $7.280$ & \cite{Eichten:1994gt} & $1^{+}$ & $7.150$ & \cite{Godfrey:2004ya} \\
		$0^{-}$ & $6.275(1)$ & \cite{ParticleDataGroup:2016lqr,McNeile:2012qf} & $0^{+}$ & $6.712$ & \cite{Mathur:2018epb} \\
		$0^{-}$ & $6.842(6)$ & \cite{ParticleDataGroup:2016lqr} & $0^{+}$ & $7.105$ & \cite{Mathur:2018epb} \\
		$0^{-}$ & $7.250$ & \cite{Godfrey:2004ya} & & & \\
		\hline\hline
	\end{tabular}	
	\caption{Relevant $B_{c}$ masses. Resonances for types $\{1^{-},1^{+},0^{-}\}$ are consistent with \cite{Bigi:2017jbd}, while resonances for $0^{+}$ are consistent with \cite{Gao:2021sav}. $B_{c} (7.280 \ \text{GeV})$ does not contribute pole to $B\to D$ process, because it is above the threshold.}
	\label{Bc_resonances}
\end{table}

Original definition of outer functions $\phi_{i}$ can be found in \cite{Boyd:1997kz}. It is convenient and universal to evaluate $\phi_{i}$ at $q^{2}=0$ where is far enough from the threshold region. All the parametrized form factors are shown as follow

\begin{align}
	f_{+}(z)&=\frac{1}{P'_{1^{-}}(z)\phi_{f_{+}}(z)}\sum^{\infty}_{n=0} b_{n}^{f_{+}}z^{n}\, ,& \quad
	f_{0}(z)&=\frac{1}{P_{0^{+}}(z)\phi_{f_{0}}(z)}\sum^{\infty}_{n=0} b_{n}^{f_{0}}z^{n}\, , \nonumber \\
	g(z)&=\frac{1}{P'_{1^{-}}(z)\phi_{g}(z)}\sum^{\infty}_{n=0} b_{n}^{g}z^{n}\,  ,& \quad
	f(z)&=\frac{1}{P_{1^{+}}(z)\phi_{f}(z)}\sum^{\infty}_{n=0} b_{n}^{f}z^{n}\, , \nonumber \\
	{\cal F}_{1}(z)&=\frac{1}{P_{1^{+}}(z)\phi_{{\cal F}_{1}}(z)}\sum^{\infty}_{n=0} b_{n}^{{\cal F}_{1}}z^{n}\,  ,&\quad
	{\cal F}_{2}(z)&=\frac{1}{P_{0^{-}}(z)\phi_{{\cal F}_{2}}(z)}\sum^{\infty}_{n=0} b_{n}^{{\cal F}_{2}}z^{n}\,.
\end{align}
Lower indices $1^{-},1^{+},0^{-},0^{+}$ means the quantum number corresponding to the type in Table~\ref{Bc_resonances}. Outer functions are evaluated at the default $N=1$

\begin{eqnarray}
	\phi_{f_{+}}(z)&=&\frac{16 r^{2}}{m_{B}}\sqrt{\frac{2 n_{I}}{3 \pi \tilde{\chi}^{T}_{1^-}(0)}}\frac{(1+z)^2 (1-z)^{1\over 2}}{[(1+r)(1-z)+2 \sqrt{r}(1+z)]^5} \, , \nonumber \\
	\phi_{f_{0}}(z)&=&2 r(1-r^{2})\sqrt{\frac{2 n_{I}}{\pi \chi^{L}_{1^-}(0)}}\frac{(1+z) (1-z)^{3\over2}}{[(1+r)(1-z)+2 \sqrt{r}(1+z)]^4} \, , \nonumber \\
	\phi_{f}(z)&=&\frac{4 r_{*}}{m_{B}^{2}}\sqrt{\frac{n_{I}}{3 \pi \chi^{T}_{1^+}(0)}}\frac{(1+z) (1-z)^{3\over2}}{[(1+r_{*})(1-z)+2 \sqrt{r_{*}}(1+z)]^4} \, , \nonumber \\
	\phi_{g}(z)&=&16 r_{*}^{2}\sqrt{\frac{n_{I}}{3 \pi \tilde{\chi}^{T}_{1^-}(0)}}\frac{(1+z)^2 (1-z)^{- \frac{1}{2}}}{[(1+r_{*})(1-z)+2 \sqrt{r_{*}}(1+z)]^4} \, , \nonumber \\
	\phi_{{\cal F}_{1}}(z)&=&\frac{4 r_{*}}{m_{B}^{3}}\sqrt{\frac{n_{I}}{6 \pi \chi^{T}_{1^+}(0)}}\frac{(1+z) (1-z)^{5\over2}}{[(1+r_{*})(1-z)+2 \sqrt{r_{*}}(1+z)]^5} \, , \nonumber \\
	\phi_{{\cal F}_{2}}(z)&=&8 r_{*}^{2}\sqrt{\frac{2 n_{I}}{\pi \tilde{\chi}^{L}_{1^+}(0)}}\frac{(1+z)^{2} (1-z)^{-\frac{1}{2}}}{[(1+r_{*})(1-z)+2 \sqrt{r_{*}}(1+z)]^4} \, , 
\end{eqnarray}
where $r=m_D/m_B,r_*=m_{D^*}/m_B$, we follow the framework established in Ref.~\cite{Bigi:2016mdz,Bigi:2017jbd}, the numerical 
values of all required input parameters are collected in Tab.~\ref{outer_function}.
\begin{table}[htbp]
	\centering
	\renewcommand{\arraystretch}{1.5}
	\begin{tabular}{cc}
		\hline\hline
		Input & Value\\
		\hline
		$n_{I}$ & $2.6$ \\
		$\tilde{\chi}^{T}_{1^-}(0)$ (GeV$^{-2}$) & $5.131 \times 10^{-4}$ \\
		$\chi^{L}_{1^-}(0)$ & $6.204 \times 10^{-3}$ \\
		$\chi^{T}_{1^+}(0)$ (GeV$^{-2}$) & $3.894 \times 10^{-4}$ \\
		$\tilde{\chi}^{L}_{1^+}(0)$ & $19.421 \times 10^{-3}$ \\
		\hline\hline
	\end{tabular}	
	\caption{Inputs for the outer functions, see Refs.~[22,39] and Ref.~\cite{Grigo:2012ji}.}
	\label{outer_function}
\end{table}

\subsection{Strong unitarity bound}\label{sec:bound}
Due to the unitarity, interaction inside $BD$ system is constrained along with those of other hadronic states with the same quantum numbers.
 These states include not only $B^{(*)}D^{(*)}$, but also $B_{c}$ resonances, a continuum of states like $ BD\pi \pi$ and etc. 
We consider $B\to D,B\to D^*,B^*\to D,B^*\to D^*$ four channels and 20 form factors in total to construct the strong unitarity bound, 
whose form factors can be studied in the heavy-quark limit \cite{Caprini:1997mu}. Short-distance corrections to the heavy quark currents 
were calculated by \cite{Neubert:1992qq} at next-to-leading order by matching from QCD onto HQET, and were collected in 
Ref.~\cite{Bernlochner:2017jka}. The second-order power correction terms in the heavy-quark effective theory for these form factors are defined by~\cite{Falk:1992wt}, including $\mathcal{O}(\varepsilon_{c}),\mathcal{O}(\varepsilon_{b}),\mathcal{O}(\varepsilon_{c}^{2}),\mathcal{O}(\varepsilon_{b}^{2})$ 
and $\mathcal{O}(\varepsilon_{c}\varepsilon_{b})$ corrections 
($\varepsilon_{c}\equiv \bar{\Lambda}/(2m_{c}),\varepsilon_{b}\equiv \bar{\Lambda}/(2m_{b})$ are small parameters, 
$\bar{\Lambda}$ is the energy carried by the light degrees of freedom in heavy meson, while we neglect the $\mathcal{O}(\varepsilon_{b}^{2})$ and $\mathcal{O}(\varepsilon_{c}\varepsilon_{b})$ 
terms. The Isgur-Wise functions, which are defined by the matrix elements of the effective operators in HQET and contribute to the power correction 
terms, are provided up to $\mathcal{O}(\varepsilon_{c}^{2})$ at TABLE~II in \cite{Bordone:2019guc} and Table~2 in \cite{Iguro:2020cpg}, 
while the former contains the effect of strange spectator quark and is preferred by us.

To obtain the form factor ratios estimated from HQET, following the form factor definition in Appendix of \cite{Caprini:1997mu}, 
we update its Table~A.1 to $\mathcal{O}(\varepsilon_{c}^{2})$ level in our Table~\ref{Rj} from Appendix~\ref{Input parameters}. We find it is 
very close comparing our results without $\mathcal{O}(\varepsilon_{c}^{2})$ 
terms to \cite{Caprini:1997mu} along with the updated one in \cite{Bigi:2017jbd}, however these corrections have non-negligible effects. 
$A_{j},B_{j},C_{j},D_{j}$ within the tables are defined by
\begin{eqnarray}
	R_{j}(\omega)\equiv\frac{F_{j}(\omega)}{V_{1}(\omega)}\equiv A_{j}\left[1+B_{j} (\omega-\omega_{0})+C_{j} (\omega-\omega_{0})^{2}+D_{j} (\omega-\omega_{0})^{3}+\dots \right] \, ,
	\label{eq_Rj}
\end{eqnarray}
where $F_{j},V_{1}$ are form factors defined in \cite{Caprini:1997mu}, and $\omega_{0}=1$ in our calculation.

In the nominal analysis the BGL series is truncated at $O(z^2)$, and the four strong unitarity inequalities in Eq.~\eqref{unitarity_bounds} are imposed as hard constraints. The coefficients entering these strong bounds are evaluated at the central HQET inputs and are kept fixed in the nominal fit. Following Ref.~\cite{Caprini:1997mu}, we use $\bar\Lambda=0.48$~GeV; we further use $m_b=4.78$~GeV and $m_c=1.67$~GeV together with the central subleading Isgur--Wise inputs of Ref.~\cite{Bordone:2019guc}. The strong bounds combine the $B\to D$, $B\to D^*$, $B^*\to D$ and $B^*\to D^*$ channels through HQET form-factor relations including the $O(\alpha_s)$, $O(\epsilon_c)$, $O(\epsilon_b)$ and $O(\epsilon_c^2)$ terms retained in our analysis. As a sensitivity test, we vary the relevant HQET inputs individually over the endpoints of their quoted ranges, reconstruct the bounds and repeat the full fit. Among 40 variations, the maximal induced shift is $0.0752\times10^{-3}$, about $0.16\sigma$ of the nominal $|V_{cb}|$ uncertainty.

We have used relations derived from equation~\eqref{fpeqf0},\eqref{F2=F1},\eqref{F1=f} to remove redundant degrees of freedom from the parameterization process,
\begin{eqnarray}
	b^{f_{0}}_{0}&=&\left.\left[P_{0^{+}}(z)\phi_{f_{0}}(z)f_{+}(z)-(b^{f_{0}}_{1}z+b^{f_{0}}_{2}z^{2})\right]\right|_{z=z(\omega_{\text{Max}},1)} \, , \nonumber \\
    b^{\mathcal{F}_2}_{0}&=&\left.\left[\frac{P_{0^{-}}(z)\phi_{\mathcal{F}_{2}}(z)(1+r_{q})}{m_{B_{q}}^{2}(1-r_{q})(1+\omega_{\text{Max}})r_{q}}\mathcal{F}_{1}(z)-b^{\mathcal{F}_2}_{1}z-b^{\mathcal{F}_2}_{2}z^{2}\right]\right|_{z=z(\omega_{\text{Max}},1)} \, , \nonumber \\
    b^{\mathcal{F}_1}_{0}&=&\frac{1-r_{q}}{\sqrt{2}\left(1+\sqrt{r_{q}}\right)^{2}} b^{f}_{0} \, .
    \label{RedundantBGLpara}
\end{eqnarray}
Then, we have the strong unitarity bound used in our fitting procedure 
\begin{align}
\label{unitarity_bounds}
1\ge &22.93\,(b_0^{g})^2 - 0.96\,b_0^{g}\, b_1^{g} +  6.49\,(b_1^{g})^2 - 5.09\,b_0^{g}\, b_2^{g}  -  3.66 \,b_1^{g}\, b_2^{g}  + 
 4.90 \,(b_2^{g})^2  
\nonumber\\
&+ (b_0^{f_+})^2 +  (b_1^{f_+})^2  + (b_2^{f_+})^2  \,,
\nonumber\\
1\ge& 17.43\,(b_0^{f})^2 + 0.47\,b_0^f\, b_1^f +  9.49\, (b_1^f)^2 - 5.56\, b_0^f\, b_2^f -  5.21\, b_1^f\, b_2^f + 5.55 \,(b_2^f)^2 + (b_1^{F_1})^2 + (b_2^{F_1})^2 \,,
\nonumber\\
1\ge & 429.14\, (b_0^{f_+})^2 + 1.80\, (b_1^{f_+})^2 + 
 0.01\, (b_2^{f_+})^2 - 0.0017\,b_2^{f_+}\, b_1^{f_0} 
+ 2.00 ( b_1^{f_0} )^2 + 0.15\, b_2^{f_+} \, b_2^{f_0}  
\nonumber\\
& +  0.05 \,b_1^{f_0} \, b_2^{f_0} + 1.95\, ( b_2^{f_0} )^2
 +   b_1^{f_+} \, (0.23 b_2^{f_+} - 0.03 \,b_1^{f_0} +  2.33 b_2^{f_0}) 
\nonumber\\
&+  b_0^{f_+}\, (55.56\, b_1^{f_+} + 3.60\, b_2^{f_+} - 0.40 \, b_1^{f_0} + 36.02\,  b_2^{f_0})\,,
\nonumber\\
1\ge & 87.20\, (b_0^f)^2 + 9.79\,(b_1^{F_1})^2 + 0.03\, (b_2^{F_1})^2 + 0.40\,b_2^{F_1}\, b_1^{F_2} +   5.02\, (b_1^{F_2})^2 
 \nonumber\\ 
&+ 0.14\,b_2^{F_1}\, b_2^{F_2}
+  3.83\,b_1^{F_2}\, b_2^{F_2} + 3.51\, (b_2^{F_2} )^2 +  b_1^{F_1}\, (1.10\, b_2^{F_1} + 7.19\, b_1^{F_2} +  2.56\, b_2^{F_2})
 \nonumber\\ 
 &+ b_0^f (58.40\, b_1^{F_1} + 3.28\, b_2^{F_1} +  21.47\, b_1^{F_2} + 7.64\, b_2^{F_2})\,.
\end{align}

We also note a limitation of the standard BGL construction emphasized recently in \cite{Gubernari:2026sqc}. Rescattering can generate subthreshold branch cuts for generic $B\to D^{(*)}$ form factors, which lie inside the unit disk of the conventional BGL map. Consequently, the standard BGL analyticity construction is, strictly speaking, approximate in these channels. A modified unitarity-bounded construction incorporating the first subthreshold cut has been developed in Ref.~\cite{Gubernari:2026sqc}; the numerical difference from plain BGL was found to be comparatively small for $B\to D^*$, although the caveat remains conceptually relevant.

\subsection{Input data points}\label{sec:input}
In this section, we introduce three categories of input data adopted in our global fits, including the theoretical predictions from LCSR calculations, 
form factor results obtained via lattice QCD simulations, and binned differential decay rate measurements reported by Belle and Belle II experiments.
\begin{itemize}
    \item \textbf{LCSR} 
The LCSR inputs used here are taken from Ref.~\cite{Cui:2023jiw}, where the $\bar B\to D^{(*)}$ form factors are calculated including NLO QCD corrections and four sources of next-to-leading-power (NLP) corrections. We use the LCSR predictions for the $\bar B\to D$ form factors $\{f_+,f_0\}$ and the $\bar B\to D^*$ form factors $\{V,A_0,A_1,A_{12}\}$ at the six representative kinematic points $q^2\in\{-3,-2,-1,0,1,2\}\,{\rm GeV}^2,$
together with their correlated uncertainties generated with a replica method (see the ancillary file \texttt{InputLCSR.txt} from the arXiv preprint version). The underlying LCSR parameters are sampled simultaneously according to the Gaussian or uniform distributions specified in Ref.~\cite{Cui:2023jiw}, and 1000 complete input-parameter sets are formed. For each set, all relevant $\bar B\to D^{(*)}$ form factors are recalculated at all six $q^2$ points. Each replica is therefore a full correlated prediction vector, so parameter variations that alter the $q^2$ dependence, including those associated with $\lambda_B$, are retained in the resulting correlations. The empirical mean vector, standard deviations, and correlation matrix of the 1000 replicas define the LCSR input used in the correlated Gaussian $\chi^2$, with the treatment of the nearly singular correlation matrix described below.

We have validated this compression of the replica ensemble in several steps. A Gaussian ensemble constructed with the same mean and covariance reproduces the point-by-point central 68\% widths within 4.5\%, while linear shape observables have the same standard deviations at the sub-percent level. The original replicas nevertheless show a more concentrated central region and heavier far tails in some directions. To test whether these non-Gaussian tails affect the fit, we rank the 1000 LCSR replicas according to their squared Mahalanobis distance
\begin{align}
	D_i^2=(\mathbf F_i-\bar{\mathbf F})^T C_{\rm emp}^{+}(\mathbf F_i-\bar{\mathbf F}),
\end{align}
where $\mathbf F_i$ denotes the complete prediction vector of the $i$-th replica, $\bar{\mathbf F}$ the empirical mean vector, and $C_{\rm emp}^{+}$ the Moore--Penrose pseudoinverse of the empirical covariance matrix $C_{\rm emp}$. The pseudoinverse is used only for this diagnostic ranking and defines the distance on the non-degenerate subspace of the strongly correlated covariance matrix. We then construct central subsets by retaining the 68\%, 84\%, and 95\% of replicas with the smallest values of $D_i^2$, recompute the LCSR mean and covariance for each subset, and repeat the complete constrained global fit; the full 100\% ensemble is the nominal case. The corresponding values of $|V_{cb}|\times10^3$ for the 68\%, 84\%, 95\%, and 100\% samples are 39.1788, 39.1826, 39.1816, and 39.1801, with uncertainties 0.4659, 0.4664, 0.4673, and 0.4681, respectively. Hence the non-Gaussian tails have a negligible effect on the present extraction.
	
For the LCSR contribution to the actual likelihood, we do not use the pseudoinverse of the raw empirical matrix. The predictions at different $q^2$ values are extremely strongly correlated, so the raw empirical correlation matrix is nearly singular and a direct inversion would assign numerically unstable large weights to directions associated with very small eigenvalues. We therefore use the mild regularization $R_{\rm reg}=\rho R_{\rm emp}+(1-\rho)I$, where $R_{\rm emp}$ is the empirical correlation matrix and $I$ is the identity matrix. Unlike the pseudoinverse used only for the Mahalanobis diagnostic, this prescription retains all correlation-matrix directions while preventing very small eigenvalues from generating unstable weights in the $\chi^2$. For $\rho=0.90$, 0.95 and 0.99 we obtain $|V_{cb}|=(39.1972,\,39.1801,\,39.1323)\times10^{-3}$ with corresponding uncertainties $(0.4688,\,0.4681,\,0.4659)\times10^{-3}$. The nominal choice $\rho=0.95$ is therefore a numerical regularization and does not materially control the extracted $|V_{cb}|$ or its uncertainty. 

Since the LFU ratios probe the form-factor shapes over the full kinematic range, we also repeat this test for the lattice-QCD+LCSR fit used to obtain our predictions for $R(D)$ and $R(D^*)$. For $\rho=0.90$, 0.95 and 0.99, respectively, we obtain
\[
\begin{array}{c|cc}
	\rho & R(D) & R(D^*)\\
	\hline
	0.90 & 0.3082\pm0.0087 & 0.2546\pm0.0048\\
	0.95 & 0.3069\pm0.0080 & 0.2548\pm0.0043\\
	0.99 & 0.3061\pm0.0070 & 0.2577\pm0.0036
\end{array}
\]
The $\rho=0.95$ values reproduce the nominal results quoted in Table~\ref{B2DandDstar}. Relative to this nominal choice, the maximal central-value shifts are $0.0012$ for $R(D)$ and $0.0029$ for $R(D^*)$, corresponding to only about $0.40\%$ and $1.13\%$ of the nominal central values, respectively. The corresponding uncertainties vary by up to approximately $13\%$ and $18\%$. Thus, the central predictions remain stable under the variation of the regularization parameter, while the uncertainties show a moderate dependence on $\rho$, particularly for $R(D^*)$. Throughout the range $\rho=0.90$--$0.99$, the predictions remain mutually compatible within their quoted uncertainties. Within the LCDA family adopted in Ref.~\cite{Cui:2023jiw}, variation of the LCDA shape parameters also probes the associated model-shape dependence, which is propagated into the LCSR covariance matrix used in our fit.
	
    \item \textbf{Lattice QCD}
We have collected recent lattice results to fix the theoretical predictions in the small recoil area: $(1)$ the lattice results for the 
$\bar{B} \to D$ form factors at $\omega=\{1.00,1.08,1.16\}$ from the FNAL/MILC Collaboration~\cite{FermilabLattice:2015ilb} and 
the synthetic data points at $\omega=\{1.01,1.06\}$ from the HPQCD analysis \cite{Na:2015kha}, $(2)$ the lattice results for the 
$\bar{B} \to D^{*}$ form factors at $\omega=\{1.03,1.10,1.17\}$ from the FNAL/MILC Collaboration \cite{FermilabLattice:2021cdg} and recent 
lattice computations at $\omega=\{1.025,1.060,1.100\}$ from JLQCD Collaboration \cite{Aoki:2023qpa}. All available correlations are considered. 
Data points by HPQCD \cite{Na:2015kha} are synthesized from six BGL parameters, thus we only keep four points because the degree of 
freedom is $5$ (constraint~$f_+(0)=f_0(0)$ decrease one degree of freedom). HPQCD \cite{Harrison:2023dzh} provides five points uniformly 
distributed over the full kinematic range $\omega=\{1.000,1.126,1.252,1.378,1.503\}$ (corresponding to 
$q^{2}=i\times (m_{B}-m_{D^{*}})^2/4,i\in[4,3,2,1,0]$), we discard the first and the last point for $\mathcal{F}_{1}$ because of 
the constraints \eqref{F1=f} and \eqref{F2=F1}, and we shield off the last two among five data points when input them together with LCSR data.

    \item \textbf{Experiments} Measurements of the binned differential decay rates $\Delta\Gamma_i/\Delta \omega$ for the semileptonic decay 
$\bar{B}\to D\ell\nu$ are available from the Belle~\cite{Glattauer:2015teq} and Belle II~\cite{Belle-II:2025rna} collaborations. The relevant 
central values, uncertainties, and correlation matrices can be found in the tables of these publications and their ancillary files. 
Partial decay rates $\Delta\Gamma$ in bins of kinematic variables, namely $\omega\,, \; \cos\theta_{\ell}\, ,\; \cos\theta_V\,,\; \chi$, 
have been measured by Belle II~\cite{Belle-II:2023okj}. By contrast, Belle provided folded data only~\cite{Belle:2018ezy}, 
one can reconstruct the binned partial decay rates by implementing the Monte Carlo (MC) unfolding procedure. 
To avoid introducing additional assumptions on the correlations between the electron and muon samples, we perform the fit using a single light-lepton channel from Ref.~\cite{Belle:2018ezy}. We choose the muon mode, for which the reconstruction is not affected by electron-specific bremsstrahlung recovery, and which also facilitates direct comparison with muon-based measurements from other experiments.

\end{itemize}

\section{Numerical results}\label{sec:result}
\begin{table}[htbp]
	\centering
	\renewcommand{\arraystretch}{1.5}
\begin{tabular}{|c|c|cccccc|}
\hline
\hline
~ & Values & \multicolumn{6}{|c|}{Correlation matrix} \\ \hline
$~|V_{cb}|$  ~&~ $\quad0.0403\pm0.0008$  ~&~1 & -0.48 & -0.31 & -0.21 & -0.23 & -0.30  \\ 
$~b_0^{f_+}$ ~&~ $\quad0.0156\pm0.0001$  ~&~ ~ &1 & 0.27 & -0.17 & 0.21 & -0.04  \\
$~b_1^{f_+}$ ~&~ $-0.0455\pm0.0037$  ~&~ ~ & ~ & 1 & -0.65 & 0.82 & -0.51  \\ 
$~b_2^{f_+}$ ~&~ $\quad0.1117\pm0.0736$  ~&~ ~ & ~ & ~ &1 & -0.56 & 0.93  \\ 
$~b_1^{f_0}$ ~&~ $-0.2281\pm0.0164$  ~&~ ~ & ~ & ~ & ~ &1 & -0.61  \\ 
$~b_2^{f_0}$ ~&~ $\quad0.5404\pm0.3428$  ~&~ ~ & ~ & ~ & ~ & ~ & 1 \\ 
\hline
\hline  
\end{tabular}
\vspace*{0.1 cm}
     \caption{Extracted values of the CKM matrix element $|V_{cb}|$  and five BGL parameters from a simultaneous fit of the BGL parameterization 
to three sets of inputs: lattice QCD simulations from Ref.~\cite{FermilabLattice:2015ilb,Na:2015kha}, correlated  SCET sum rule predictions 
at large hadronic recoil~\cite{Cui:2023jiw}, and binned differential $B\to D\ell\nu$ decay spectra~\cite{Glattauer:2015teq,Belle-II:2025rna}.  
Correlations between the fitting parameters are also provided.}
	\label{B2Donly}
\end{table}

We perform a simultaneous global fit of the BGL form factor parameterization for the exclusive semileptonic decay $B\to D\ell\nu$ using three
 complementary categories of theoretical and experimental inputs:  lattice QCD simulations from  Ref.~\cite{FermilabLattice:2015ilb,Na:2015kha},
 correlated large hadronic recoil SCET sum rule predictions given in Ref.~\cite{Cui:2023jiw}, and binned differential decay spectra measured in
 Ref.~\cite{Glattauer:2015teq,Belle-II:2025rna}. The full set of fitted quantities includes the CKM matrix element $|V_{cb}|$  together with 
five BGL shape coefficients, namely three parameters $b_0^{f_+},b_1^{f_+},b_2^{f_+}$ describing the $f_+$ form factor and two parameters
 $b_1^{f_0},b_2^{f_0}$ for the $f_0$ form factor; the central values and associated fit uncertainties of all six fitted parameters are
 summarized in the second column of Table.~\ref{B2Donly}. The extracted result for the CKM matrix element reads $|V_{cb}|=(40.3\pm0.8)\times 10^{-3}$.
Strong linear correlations exist among these fitting parameters, which are fully documented in the correlation matrix displayed on 
the right-hand side of Table.~\ref{B2Donly}. For instance, the BGL coefficient $b_2^{f_+}$ is highly positively correlated with 
$b_2^{f_0}$, with a correlation coefficient of 0.93; meanwhile, $|V_{cb}|$  carries moderate negative correlations with all five BGL shape parameters, with 
correlation coefficients ranging from $-0.21$ to $-0.48$. These nontrivial inter-parameter 
correlations must be fully retained during any subsequent uncertainty propagation for physical observables  derived from the fitted \(|V_{cb}|\) 
and the complete set of BGL form-factor coefficients, as neglecting such couplings would lead to underestimated or biased total uncertainties for 
all downstream predictions. We first compare our fit with that of Ref.~\cite{Gao:2021sav}. Both analyses employ the model-independent BGL $z$-parameterization scheme to describe the 
$B\to D$ semileptonic transition form factors $f_+(q^2)$ and $f_0(q^2)$, as well as extract the CKM matrix element $|V_{cb}|$. 
The smaller uncertainty of $|V_{cb}|$ obtained from our fit may be attributed to several factors: we incorporate the experimental 
measurements released by Belle II~\cite{Belle-II:2025rna}, account for the correlations between distinct data points of the 
LCSR predictions in the small-$q^2$ region, and include higher-order terms of the BGL expansion. In addition, the strong unitarity bounds 
adopted in our analysis also contain higher-power expansion terms, which exert minor influences on the fitted results. Unfortunately, 
Ref.~\cite{Fang:2026hru} does not provide fitted values of $|V_{cb}|$ obtained from Scenario A using only the $B\to D\ell\nu$ channel alone.
Nevertheless, when comparing our results with those extracted under scenario B from Ref.~\cite{Fang:2026hru}, we find that 
our uncertainties on $|V_{cb}|$ are identical to those reported in the literature, while our central value is larger by $3.3\%$.

\begin{table}[t]
	\centering
	\renewcommand{\arraystretch}{1.5}
\begin{tabular}{|c|c|ccccccccccc|}
\hline
\hline
~ & Values & \multicolumn{11}{|c|}{Correlation matrix} \\ \hline
$|V_{cb}|$  & $\quad0.0386\pm0.0006$  & 1 & -0.22 & -0.09 & -0.01 & -0.61 & -0.33 & 0.04 & -0.33 & 0.16 & -0.25 & 0.18 \\
$b_0^g$ & $\quad0.0270\pm0.0006$ & ~ & 1 & -0.22 & -0.16 & 0.29 & 0.02 & -0.07 & 0.13 & -0.07 & 0.01 & -0.06 \\
$b_1^g$ & $-0.0666\pm0.0243$ & ~ & ~ & 1 & -0.76 & 0.01 & 0.20 & -0.19 & 0.12 & -0.07 & 0.21 & -0.18 \\
$b_2^g$ & $-0.4354\pm0.5901$ & ~ & ~ & ~ & 1 & 0.03 & -0.15 & 0.18 & -0.05 & -0.02 & -0.13 & 0.11 \\
$b_0^f$ & $\quad0.0122\pm0.0001$ & & ~ & ~ & ~ & 1 & -0.04 & 0.06 & -0.02 & 0.00 & 0.05 & -0.09 \\
$b_1^f$ & $\quad0.0149\pm0.0056$ & & ~ & ~ & ~ & ~ & 1 & -0.79 & 0.57 & -0.42 & 0.44 & -0.36 \\
$b_2^f$ & $-0.2624\pm0.1225$ & & ~ & ~ & ~ & ~ & ~ & 1 & -0.39 & 0.30 & -0.34 & 0.30 \\
$b_1^{F_1}$ & $\quad0.0019\pm0.0011$ & & ~ & ~ & ~ & ~ & ~ & ~ & 1 & -0.93 & 0.46 & -0.55 \\
$b_2^{F_1}$ & $-0.0391\pm0.0195$ & & ~ & ~ & ~ & ~ &~ &~ & ~ & 1 & -0.38 & 0.56 \\
$b_1^{F_2}$ & $-0.1144\pm0.0348$ & & ~ & ~ & ~ & ~ & ~ & ~ & ~ & ~ & 1 & -0.89 \\
$b_2^{F_2}$ & $-0.2345\pm0.5779$ &  & ~ & ~ & ~ & ~ & ~ & ~ & ~ & ~ & ~ & 1 \\
\hline
\hline
\end{tabular}
\vspace*{0.1 cm}
     \caption{Extracted values of the CKM matrix element $|V_{cb}|$ and ten independent BGL parameters from a simultaneous fit of the BGL parameterization 
to three sets of inputs: lattice QCD simulations from Ref.~\cite{FermilabLattice:2021cdg,Harrison:2023dzh,Aoki:2023qpa}, 
correlated  SCET sum rule predictions at large hadronic recoil~\cite{Cui:2023jiw}, and binned differential $B\to D^*\ell\nu$ decay 
spectra~\cite{Belle:2018ezy,Belle-II:2023okj}. Correlations between the fitting parameters are also provided.}
	\label{B2Dstaronly}
\end{table}
Along the same vein, we then perform a simultaneous global fit of the BGL dispersive parameterization dedicated to the semileptonic 
\(\bar{B}\to D^*\ell \nu \) transition by incorporating three distinct categories of theoretical and experimental constraints, which include 
lattice QCD form factor simulations detailed in Refs.~\cite{FermilabLattice:2021cdg,Harrison:2023dzh,Aoki:2023qpa}, correlated large hadronic 
recoil SCET sum rule predictions from Ref.~\cite{Cui:2023jiw}, and binned differential decay spectra measured for the $B\to D^*\ell\nu$ channel as 
reported in Belle~\cite{Belle:2018ezy} and Belle II~\cite{Belle-II:2023okj}; the full set of fitted quantities extracted from this global minimization 
consists of the CKM matrix element 
$|V_{cb}|$  alongside ten independent BGL coefficients, the central values and associated fit uncertainties of all eleven fitted parameters 
are summarized in Table.~\ref{B2Dstaronly}. All pairs of fitted parameters show non-negligible linear correlations, and the 
full correlation matrix covering every pairwise correlation coefficient is displayed on the right side of this table. Strong correlations arise 
between BGL coefficients belonging to a single form factor; for example, \(b_1^{F_1}\) and \(b_2^{F_1}\) share a strong negative correlation of 
\(-0.93\), while \(b_1^{F_2}\) and \(b_2^{F_2}\) are negatively correlated at \(-0.89\). For the CKM matrix element $|V_{cb}|$, moderate negative 
correlations exist with nearly all BGL shape coefficients, with correlation coefficients spanning \(-0.61\) to \(-0.01\), alongside mild positive 
correlations with a small number of coefficients. Any subsequent uncertainty propagation for physical observables calculated using the fitted 
\(|V_{cb}|\) and all BGL form-factor coefficients requires full inclusion of these meaningful parameter correlations. Ignoring these mutual couplings 
may produce biased or underestimated total uncertainties for all resulting physical predictions. Our extracted value 
\(|V_{cb}|=(38.6\pm0.6)\times 10^{-3}\) falls about \(1.7\times 10^{-3}\) below the Type-B combined result \((40.29\pm0.71)\times 10^{-3}\)
of Ref.~\cite{Bordone:2026gqs}, even though both analyses employ a simultaneous fit of BGL coefficients and \(|V_{cb}|\) to lattice and
experimental data, tangible disparities in extracted \(|V_{cb}|\) values arise from four key methodological distinctions embedded in our 
input setup and statistical paradigm: we incorporate correlated SCET sum rule predictions as supplementary theoretical constraints absent from the 
reference's fit pool, remove electron-channel measurements from the Belle18~\cite{Belle:2018ezy} dataset to mitigate well-documented tension between 
electron and muon spectra, discard only normalized differential decay distributions provided in Ref.~\cite{Belle:2023xgj}, and exclusively employ 
full unnormalized angular observables tabulated in \cite{Belle-II:2023okj}, and adopt a constrained frequentist $\chi^2$ minimization, whereas Ref.~\cite{Bordone:2026gqs} performs the Bayesian joint analysis using nested sampling implemented with \texttt{PyMultiNest}; the combined effect of richer correlated theory inputs, tension-mitigated experimental 
subsets, and fundamentally different statistical treatment of  parameter uncertainties jointly shifts the central value and narrows the total 
uncertainty of our \(|V_{cb}|\) result relative to the Bayesian extraction reported in Ref.~\cite{Bordone:2026gqs}. Fully differential angular observables in $\bar B\to D^*\ell\bar\nu$ provide complementary information on the hadronic form factors and on possible non-standard semileptonic interactions; see Refs.~\cite{Colangelo:2018cnj,Colangelo:2024mxe} for related phenomenological studies.
It is also interesting to see that our extracted central value of \(|V_{cb}|\) is lower by \(1.2\times10^{-3}\) yet yields 
a comparable uncertainty to the \((39.8\pm0.6)\times10^{-3}\) result from the fit scenario of \((B)+\mathcal{B}(B^+\to \bar{D}^{0*}\ell^+\nu)\) 
channel presented in Ref.~\cite{Fang:2026hru}, and this central value offset originates from multiple intertwined methodological distinctions.
Ref.~\cite{Fang:2026hru} also retains the LCSR correlations in its correlated-fit analysis, as in our nominal fit. We do not include the Belle 2017 dataset~\cite{Belle:2017rcc}, preferring to base the nominal fit on the more recent experimental inputs, whereas Ref.~\cite{Fang:2026hru} retains this measurement. The two analyses adopt different free-parameter 
setups in the global fit with divergent treatments of \(|V_{cb}|\) as a fitted observable, and more fundamentally, the reference relies on 
Bayesian MCMC sampling whereas our work implements frequentist maximum-likelihood minimization to extract form-factor and CKM parameters 
simultaneously, with all these differences collectively shifting our fitted \(|V_{cb}|\) central value downward without noticeably altering the 
total uncertainty budget.

\begin{table}[htbp]
	\centering
	\renewcommand{\arraystretch}{1.5}
\begin{tabular}{|c|ccc|}
\hline
\hline
~ & \multicolumn{3}{|c|}{Fit scenario} \\ 
\hline
 Values  ~&~ Lattice ~&~ Lattice $\oplus$ LCSR  ~&~ Lattice $\oplus$ LCSR   $\oplus$ Exp.  \\
\hline
$|V_{cb}|$  ~&~ -  ~&~  -  ~&~ $(39.18\pm0.47)\times10^{-3}$\\ 
\hline
$b_0^{f_+}$ ~&~ $\quad0.0157\pm0.0001$  ~&~ $\quad0.0156\pm0.0001$ ~&~ $\quad0.0157\pm 0.0001$\\
$b_1^{f_+}$ ~&~ $-0.0470\pm0.0038$  ~&~ $-0.0476\pm 0.0037$ ~&~ $-0.0436\pm  0.0035$\\ 
$b_2^{f_+}$ ~&~ $\quad0.1145\pm0.2288$  ~&~ $\quad0.0444\pm0.1264$ ~&~ $\quad0.1036\pm  0.0734$\\ 
$b_1^{f_0}$ ~&~ $-0.2345\pm0.0177$  ~&~ $-0.2365\pm0.0167$  ~&~ $-0.2211\pm  0.0162$\\
$b_2^{f_0}$ ~&~ $\quad0.5394\pm1.0975$  ~&~ $\quad0.1989\pm 0.5989$ ~&~  $\quad0.5409\pm  0.3356$\\ 
\hline
$b_0^g$ ~&~ $\quad0.0279\pm0.0007$ ~&~ $\quad0.0279\pm0.0007$ ~&~ $\quad0.0269\pm  0.0006$ \\
$b_1^g$ ~&~ $-0.0433\pm0.0322$ ~&~ $-0.0424\pm0.0265$ ~&~ $-0.0679\pm  0.0242$\\
$b_2^g$ ~&~ $-0.4426\pm1.5919$ ~&~ $-0.1844\pm0.6225$ ~&~ $-0.4490\pm 0.5884$\\
$b_0^f$ ~&~ $\quad0.0121\pm0.0001$ ~&~ $\quad0.0121\pm0.0001$ ~&~ $\quad0.0121\pm  0.0001$ \\
$b_1^f$ ~&~ $\quad0.0130\pm0.0068$ ~&~ $\quad0.0134\pm 0.0065$ ~&~ $\quad0.0130\pm 0.0055$\\
$b_2^f$ ~&~ $-0.1310\pm0.3011$ ~&~ $\quad0.0852\pm0.1720$ ~&~ $-0.2604\pm  0.1219$\\
$b_1^{F_1}$ ~&~ $\quad0.0034\pm0.0017$ ~&~ $-0.0030\pm0.0016$ ~&~ $\quad0.0015\pm  0.0010$\\
$b_2^{F_1}$ ~&~ $-0.0361\pm0.0616$ ~&~ $\quad0.0144\pm0.0350$ ~&~  $-0.0350\pm  0.0192$\\
$b_1^{F_2}$ ~&~ $-0.1913\pm0.0428$ ~&~ $-0.1809\pm0.0412$ ~&~ $-0.1215\pm  0.0343$\\
$b_2^{F_2}$ ~&~ $-0.4075\pm1.2661$ ~&~ $\quad0.5909\pm0.7470$ ~&~   $-0.1539\pm  0.5720$\\
\hline
$R(D)$ ~&~ $\quad0.3018\pm0.0144$ ~&~ $\quad0.3069\pm0.0080$  ~&~ $\quad0.2983\pm0.0034$  \\
$R(D^*)$ ~&~ $\quad0.2628\pm0.0082$  ~&~ $\quad0.2548\pm0.0043$ ~&~ $\quad0.2538\pm0.0011$ \\
\hline
\hline  
\end{tabular}
\vspace*{0.1 cm}
     \caption{Fitted central values and uncertainties of the full set of BGL expansion coefficients for \(\bar{B}\to D^{(*)}\) semileptonic 
form factors, extracted under three distinct fit scenarios. The first column corresponds to the fit 
constrained solely by lattice QCD form factor 
predictions~\cite{FermilabLattice:2015ilb,Na:2015kha,FermilabLattice:2021cdg,Harrison:2023dzh,Aoki:2023qpa}; the second column presents 
joint fitting results combining lattice QCD and correlated SCET sum rule theoretical inputs~\cite{Cui:2023jiw}; the final column displays 
the complete global fit results incorporating lattice QCD, LCSR calculations, and experimental data 
points~\cite{Glattauer:2015teq,Belle-II:2025rna,Belle:2018ezy,Belle-II:2023okj}. The CKM matrix element \(|V_{cb}|\) is only extracted in the 
full global fit with experimental data included. Correlation matrices of the parameters for the three fitting schemes are presented in 
Appendix~\ref{Correlation matrix}. }
	\label{B2DandDstar}
\end{table}

We then carry out three separate fits of the BGL dispersive parameterization for semileptonic \(\bar{B}\to D^{(*)}\) form factors with 
progressively enlarged constraint sets, and the extracted central values together with fit uncertainties of all 
BGL expansion coefficients are summarized in Table~\ref{B2DandDstar}. The first fit scenario relies purely on lattice QCD 
form factor simulations from Refs.~\cite{FermilabLattice:2015ilb,Na:2015kha,FermilabLattice:2021cdg,Harrison:2023dzh,Aoki:2023qpa}. 
The second analysis jointly incorporates lattice QCD data and correlated SCET sum rule predictions detailed in Ref.~\cite{Cui:2023jiw}.
The first two fit scenarios rely solely on theoretical inputs without experimental decay data, so \(|V_{cb}|\) remains undetermined 
in these theory-only fits. The final global fit supplements the lattice-QCD and LCSR inputs with measured binned differential $\bar B\to D^{(*)}\ell\bar\nu$ decay distributions from Belle and Belle~II \cite{Glattauer:2015teq,Belle-II:2025rna,Belle:2018ezy,Belle-II:2023okj}, which enables us to determine the BGL coefficients and the CKM matrix element $|V_{cb}|$ simultaneously. The numerical analysis is performed by minimizing a correlated Gaussian $\chi^2$,
\begin{align}
	\chi^2=\chi^2_{\rm exp}+\chi^2_{\rm lat}+\chi^2_{\rm LCSR},\qquad
	\chi^2_k=\Delta_k^T C_k^{-1}\Delta_k,
\end{align}
where $k\in\{\mathrm{exp},\mathrm{lat},\mathrm{LCSR}\}$ labels the three input classes, $\Delta_k(\boldsymbol\theta)\equiv T_k(\boldsymbol\theta)-D_k$ is the residual vector between the theory prediction $T_k$ and the corresponding input vector $D_k$, and $C_k$ is the covariance matrix of that input set. The full covariance information available for each dataset is retained, and the four strong unitarity inequalities in Eq.~(18) are imposed as hard constraints. For the complete lattice$\oplus$LCSR$\oplus$experiment fit we obtain $\chi^2_{\rm min}=218.505$ for 229 degrees of freedom, corresponding to a $p$-value of 0.680. The lattice-QCD, LCSR and experimental contributions are 42.91, 14.44 and 161.16, respectively. At the minimum the four strong-bound quadratic forms are $0.9999991$, $0.4170254$, $0.9999999$ and $0.1960180$ in the order of the $1^-$, $1^+$, $0^+$ and $0^-$ channels.
Comparing across the three columns, 
noticeable shifts in the central values and considerable shrinkage of uncertainties for many high-order BGL coefficients can be observed 
after introducing LCSR theoretical constraints and further experimental data, as additional observables efficiently narrow 
the allowed parameter space of the form-factor expansion. 
Noticeable shifts appear in the fitted central values of BGL coefficients and the extracted \(|V_{cb}|\), accompanied by 
an overall reduction of total uncertainties compared to Ref.~\cite{Cui:2023jiw}. These changes arise exclusively from the two 
newly added input datasets, the supplementary \(B\to D^*\) lattice form factor calculations from HPQCD~\cite{Harrison:2023dzh} and 
JLQCD~\cite{Aoki:2023qpa} tighten the theoretical bounds on the nonperturbative shape parameters, while the additional 
Belle II~\cite{Belle-II:2025rna,Belle-II:2023okj} experimental measurements introduce more precise kinematic decay observables 
that further constrain the parameter space. All other theoretical inputs, experimental samples and the fitting procedure remain identical to 
the original analysis~\cite{Cui:2023jiw}, so the observed central value deviations and narrower error bands can be fully attributed to 
these two extended data sources. Our extracted \(|V_{cb}|=(39.18\pm0.47)\times10^{-3}\) exhibits a comparable uncertainty magnitude yet 
a moderately lower central value relative to the \((39.5\pm0.5)\times10^{-3}\) outcome from the combined \(\bar{B}\to D\) and \(\bar{B}\to D^{*}\) 
(B)+ BGL (\(N=2\)) combined scenario fit presented in Ref.~\cite{Fang:2026hru}, and this mild central value discrepancy arises from 
multiple systematic differences in input selection and fitting methodology. Ref.~\cite{Fang:2026hru} also retains the LCSR correlations in its correlated-fit analysis, as in our nominal fit. The two analyses nevertheless differ in their input selections and statistical inference frameworks. We deliberately exclude the Belle17 dataset~\cite{Belle:2017rcc} entirely and only retain the muonic mode component of 
Belle18 measurements~\cite{Belle:2018ezy}, whereas the cited work adopts the full set of these older experimental inputs without screening. 
Divergent treatments of the CKM matrix element \(|V_{cb}|\) as a free fitted parameter further separate the two analyses, alongside fundamentally 
distinct statistical inference pipelines. Ref.~\cite{Fang:2026hru} employs Bayesian MCMC sampling, whereas the present analysis determines the parameters by constrained frequentist $\chi^2$ minimization, with all these combined factors driving the downward shift of our fitted \(|V_{cb}|\) central value without substantial change to 
the overall uncertainty budget.

The parameter correlation matrix quoted in Table~\ref{B2DandDstar} is obtained from the inverse Hessian of $\chi^2/2$ at the constrained minimum. As a check of the effect of the active unitarity boundaries, a constrained profile-likelihood scan gives $|V_{cb}|=(39.1801^{+0.4611}_{-0.4577})\times10^{-3}$, consistent with the Hessian uncertainty $0.4681\times10^{-3}$. We also repeat the complete fit after increasing the BGL truncation from $O(z^2)$ to $O(z^3)$. This introduces six additional cubic coefficients and lowers the minimum $\chi^2$ only from $218.505$ to $213.853$, i.e. a decrease of only $4.65$ for six additional parameters. The newly introduced coefficients have very large uncertainties and generate nearly flat directions in parameter space, while all four strong unitarity bounds become essentially saturated at the $O(z^3)$ minimum. This shows that the present data constrain the additional cubic coefficients only weakly: the extra freedom is largely absorbed along flat directions until limited by the unitarity boundaries, rather than being directly determined by the data. Nevertheless, the extracted CKM element is very stable. The central value changes from $39.1801\times10^{-3}$ to $39.2121\times10^{-3}$, a shift of only $0.0320\times10^{-3}$, and the constrained profile interval becomes $|V_{cb}|=(39.2121^{+0.4633}_{-0.4576})\times10^{-3}$, essentially unchanged from the nominal result.

\begin{figure}[htbp]
\begin{center}
\includegraphics[width=0.45 \columnwidth]{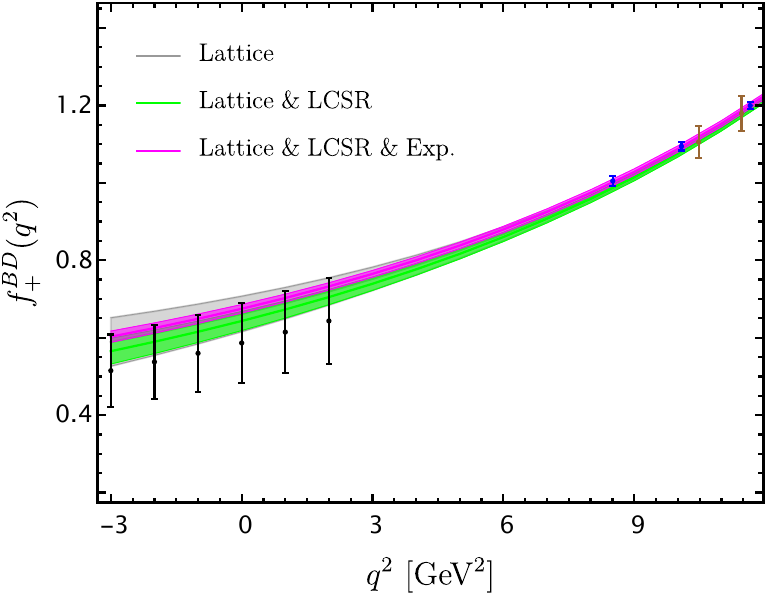}
\hspace{1.0 cm}
\includegraphics[width=0.45 \columnwidth]{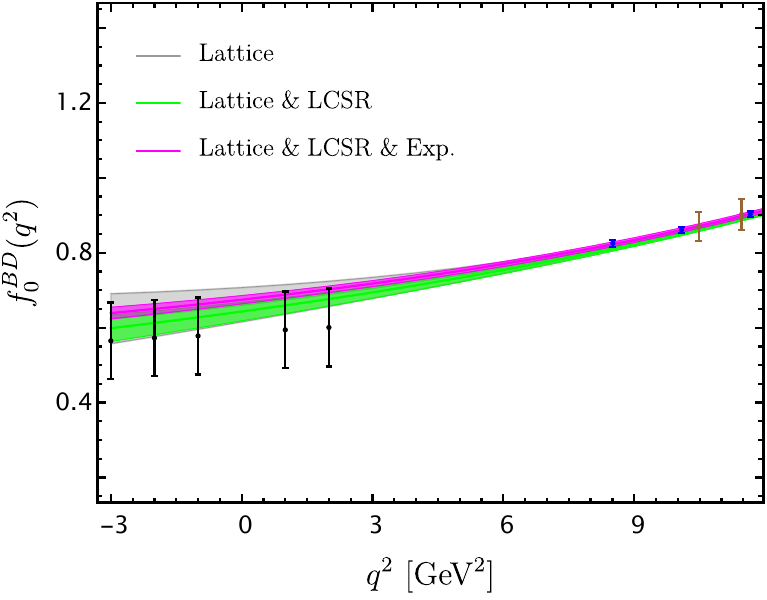}
\vspace*{0.1 cm}
\caption{Theory predictions for the momentum transfer dependence of the complete set of the exclusive
$B\to D\ell\nu$ ​  form  factors in the entire kinematic region from: 
I) the BGL $z$-series fit against the only lattice QCD data points~\cite{FermilabLattice:2015ilb,Na:2015kha},
II) the simultaneous fit to both the lattice QCD results~\cite{FermilabLattice:2015ilb,Na:2015kha} and SCET sum rule predictions~\cite{Cui:2023jiw}, 
III) the combined numerical fit including further the available experimental data 
points~\cite{Glattauer:2015teq,Belle-II:2025rna}.}
\label{fig:q2depB2DFF}
\end{center}
\end{figure}
We observe from Fig.~\ref{fig:q2depB2DFF} that the form factors $f_+^{BD}​(q^2)$ and $f_0^{BD}​(q^2)$ extracted from three successive BGL 
$z$-series fitting strategies exhibit mild but systematic shifts across the full kinematic range of $q^2$, where the lattice-only fit (grey band) 
serves as the baseline constrained solely by lattice QCD inputs, the combined lattice and LCSR fit (green band) incorporates SCET sum-rule predictions 
at large hadronic recoil and visibly tightens the theoretical uncertainty envelope especially in the small $q^2$ region, and the global 
lattice, LCSR and experiment fit (magenta band) further modifies the central values while slightly reducing uncertainties by embedding 
experimental decay data. The black error bars corresponding to SCET sum-rule predictions populate the small-$q^2$ regime and provide crucial 
constraints to regularise the $z$-series expansion away from the zero-recoil point, the blue error bars denote the MILC~\cite{FermilabLattice:2015ilb} 
lattice QCD data covering large momentum transfers, and the brown error bars represent the recent HPQCD~\cite{Na:2015kha} lattice results 
at large $q^2$, with consistent overlap visible among all three sets of discrete theoretical inputs, mild tension can be identified between 
lattice QCD results and LCSR/SCET predictions, even though the data generally overlap within their uncertainty bands. 
The global lattice, LCSR and experiment fit (magenta band)  introduces a modest central value shift relative to the lattice and LCSR theoretical fit 
and achieves only a marginal reduction of theoretical uncertainties after embedding experimental decay data.

\begin{figure}[htbp]
\begin{center}
\includegraphics[width=0.45 \columnwidth]{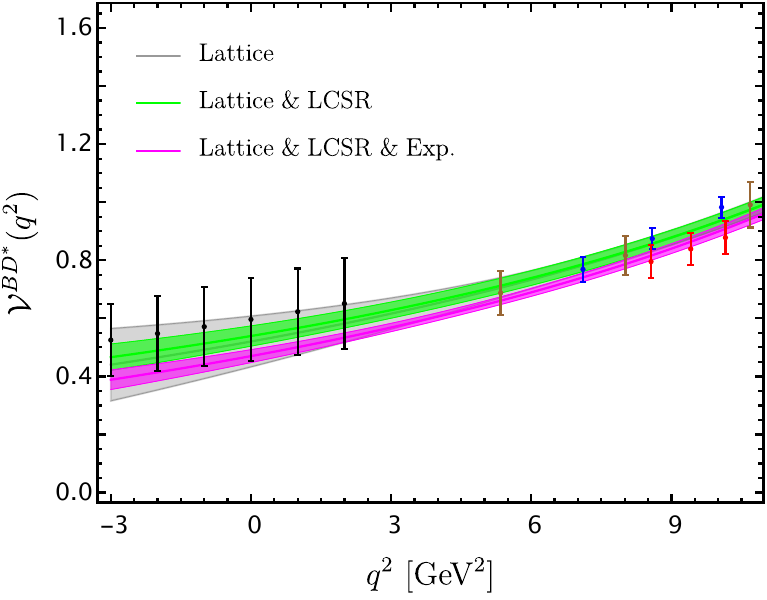}
\hspace{1.0 cm}
\includegraphics[width=0.45 \columnwidth]{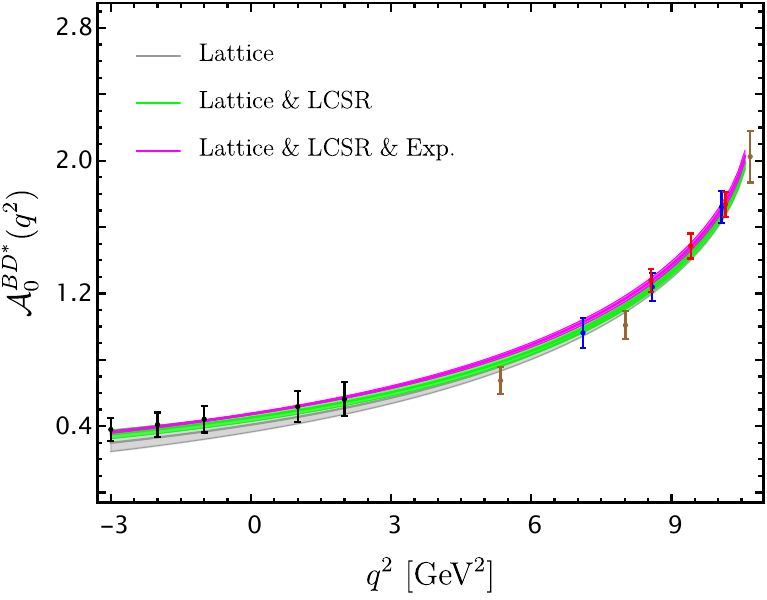}
\\
\vspace*{0.2 cm}
\includegraphics[width=0.45 \columnwidth]{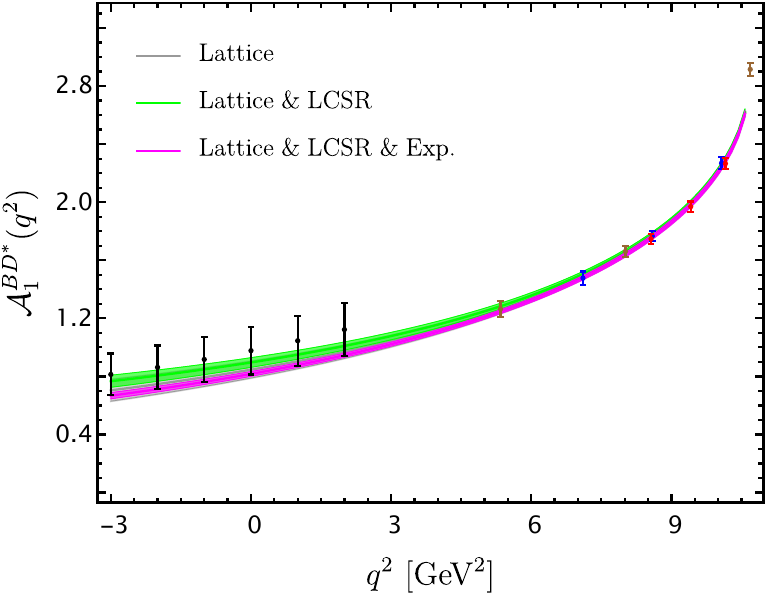}
\hspace{1.0 cm}
\includegraphics[width=0.45 \columnwidth]{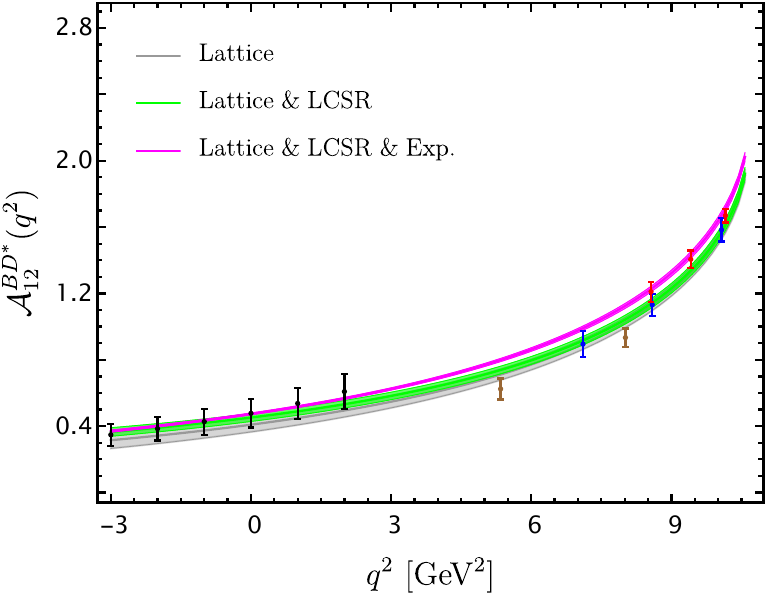}
\vspace*{0.1 cm}
\caption{Theory predictions for the momentum transfer dependence of the complete set of the exclusive $B \to D^* \ell\nu$  form  factors 
in the entire kinematic region from: 
I) the BGL $z$-series fit against the only lattice QCD data points~\cite{FermilabLattice:2021cdg,Harrison:2023dzh,Aoki:2023qpa}, 
II) the simultaneous fit to both the lattice QCD results~\cite{FermilabLattice:2021cdg,Harrison:2023dzh,Aoki:2023qpa} and SCET sum rule 
predictions~\cite{Cui:2023jiw},
III) the combined numerical fit including further the available experimental data points~\cite{Belle:2018ezy,Belle-II:2023okj}.}
\label{fig:q2depB2DstarFF}
\end{center}
\end{figure}
It can be seen from Fig.~\ref{fig:q2depB2DstarFF} that the $B\to D^*\ell\nu$ form factors ${\mathcal V}^{BD^*}(q^2)$, 
${\mathcal A}_ 0^{BD^*}(q^2)$, ${\mathcal A}_ 1^{BD^*}(q^2)$ and ${\mathcal A}_ {12}^{BD^*}(q^2)$ (detailed definition of above $B\to D^*$ form factors
can be found in Ref.~\cite{Cui:2023jiw}) obtained via three nested 
BGL $z$-series fitting strategies covering the full physical kinematic range of 
momentum transfer. The lattice-only fit (grey uncertainty band) constrained purely by lattice QCD datasets, the combined 
lattice and LCSR fit (green band) incorporates SCET sum-rule inputs at large hadronic recoil and effectively narrows theoretical uncertainties in the
small $q^2$ region, while the global lattice, LCSR and experiment fit (magenta band) further adjusts the central values and moderately reduces 
uncertainties by incorporating experimental measurements. The black error bars denote SCET sum-rule predictions that provide essential small $q^2$ 
constraints to stabilise the $z$-series extrapolation far from the zero-recoil point, blue error bars correspond to the 
MILC~\cite{FermilabLattice:2021cdg} lattice QCD results, brown error bars represent HPQCD~\cite{Harrison:2023dzh} lattice data, and red error bars 
stand for the JLQCD~\cite{Aoki:2023qpa} lattice calculations at large $q^2$. It is evident that visible tension can be identified between 
the HPQCD dataset and the remaining lattice QCD results, and mild tension also exists between LCSR predictions and lattice inputs, 
though most data points still overlap within their quoted uncertainty intervals. One can observe that the relatively small separation between 
the green and magenta bands demonstrates that experimental observables induce mild corrections to the form-factor shapes primarily determined 
by lattice QCD and SCET sum rules.

\begin{figure}[htbp]
\begin{center}
\includegraphics[width=0.8 \columnwidth]{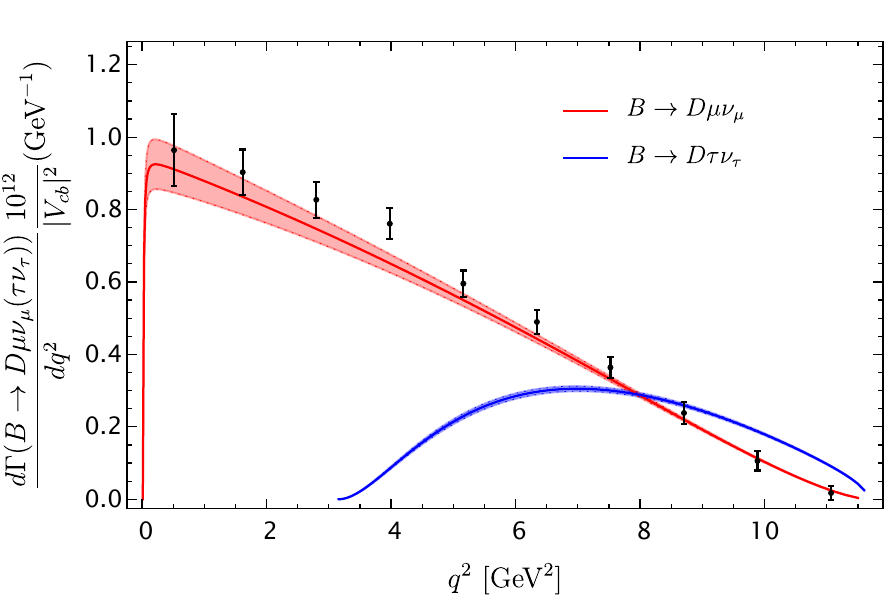}
\vspace*{0.1 cm}
\caption{Theoretical predictions (Lattice and LCSR joint fit scenario from Table.~\ref{B2DandDstar}) for the differential decay rates 
$B\to D\mu\nu_{\mu}$ (red band) and  $B\to D\tau\nu_{\tau}$ (blue band) as a function of the momentum transfer squared $q^2$. 
For comparison, the Belle II experimental data points~\cite{Belle-II:2025rna} (black error bars) for neutral semileptonic decays 
$B^0\to D^-\mu^+\nu_{\mu}$ are also presented (we use \(|V_{cb}|=39.18\times 10^{-3}\) from Table.~\ref{B2DandDstar}).}
\label{fig:diffB2D}
\end{center}
\end{figure}
To achieve a more intuitive comparison between the theoretical predictions for the differential width obtained via the SCET sum rule combined 
with lattice QCD and the corresponding experimental measurements, we then show in Fig.~\ref{fig:diffB2D} the theoretical predictions for 
the differential decay width $d\Gamma/dq^2$ of semileptonic decays
$B\to D\mu\nu_{\mu}$ and $B\to D\tau\nu_{\tau}$​, obtained from the joint lattice QCD and LCSR fit scenario in Table.~\ref{B2DandDstar}. 
The red shaded band corresponds  to $B\to D\mu\nu_{\mu}$​, while the blue band denotes $B\to D\tau\nu_{\tau}$​. 
Black vertical error bars represent the Belle II measurements~\cite{Belle-II:2025rna} 
for $B^0\to D^-\mu^+\nu_{\mu}$​. The spectrum of the muon channel rises sharply near $q^2\to 0$ and decreases monotonically over the full 
kinematic range. Owing to the larger mass of the tauon, the $B\to D\tau\nu_{\tau}$​ process has a nonvanishing kinematic threshold at finite 
$q^2$. The Belle II experimental data points~\cite{Belle-II:2025rna} follow the overall trend of the theoretical band for $B\to D\mu\nu_{\mu}$​. 
Within current uncertainties, the experimental measurements are consistent with the lattice and LCSR theoretical expectation.

\begin{figure}[htbp]
\begin{center}
\includegraphics[width=0.45 \columnwidth]{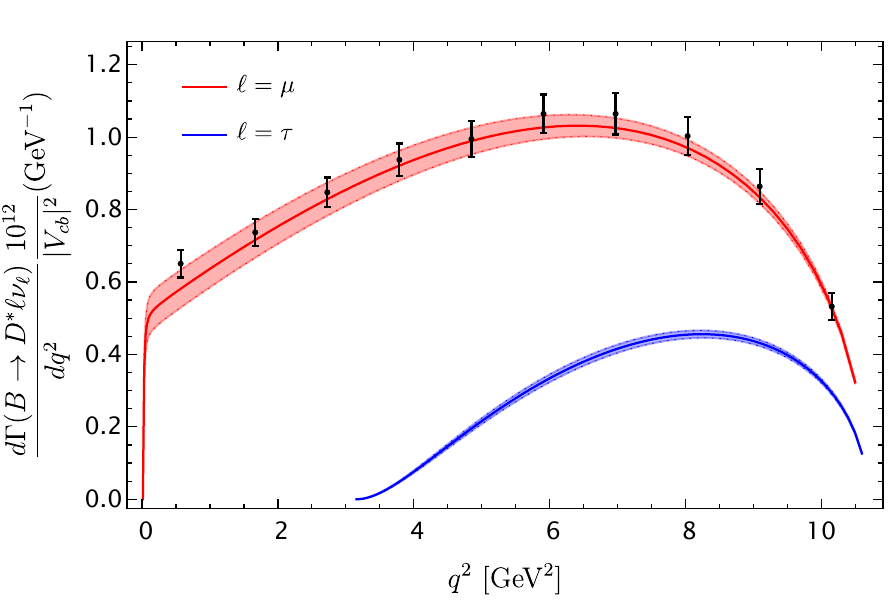}
\hspace{1.0 cm}
\includegraphics[width=0.45 \columnwidth]{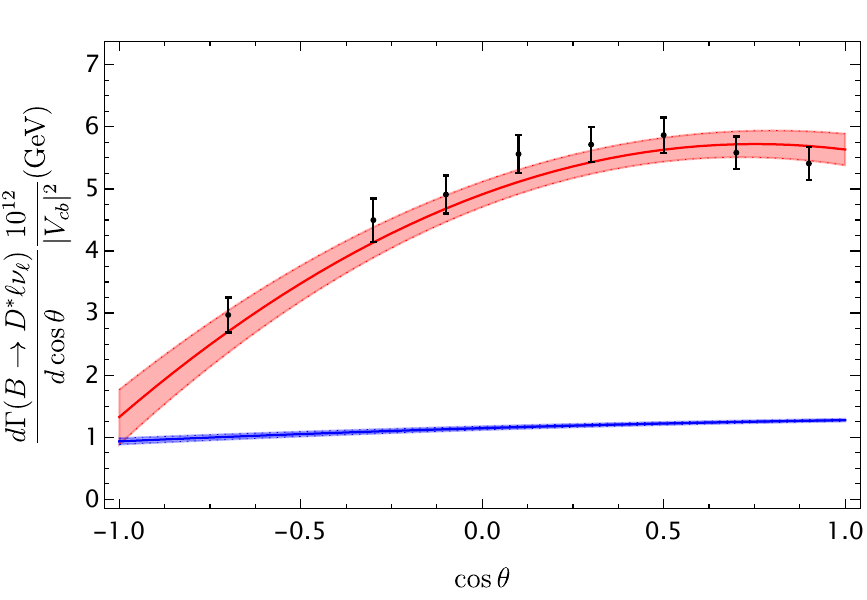}
\\
\vspace*{0.2 cm}
\includegraphics[width=0.45 \columnwidth]{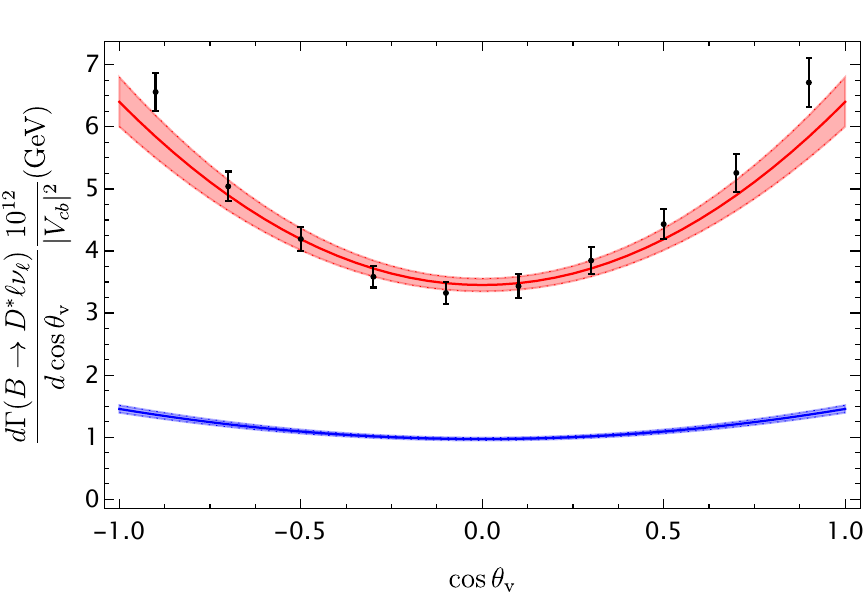}
\hspace{1.0 cm}
\includegraphics[width=0.45 \columnwidth]{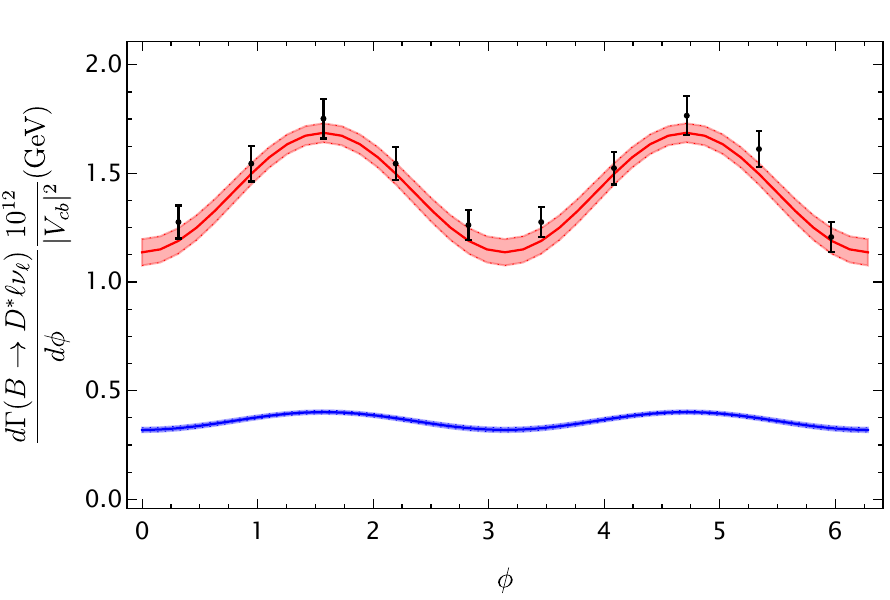}
\vspace*{0.1 cm}
\caption{The theoretical predictions (lattice and LCSR fit scenario from Table.~\ref{B2DandDstar}) for the differential decay width 
$d\Gamma(B\to D^*\ell\nu_\ell)/dX$ as functions of $q^2$, $\cos\theta$, $\cos\theta_v$ and $\phi$, shown in the four panels from top-left, top-right, 
bottom-left to bottom-right, respectively. Red bands correspond to $\ell=\mu$ and blue curves denote $\ell=\tau$. Black vertical error bars 
represent binned experimental measurements from Belle~II~\cite{Belle-II:2023okj} (we use \(|V_{cb}|=39.18\times 10^{-3}\) from Table.~\ref{B2DandDstar}). 
The shape differences between muonic and tauonic distributions 
originate from the lepton-mass effects in the semileptonic $B\to D^*\ell\nu_\ell$ decays within the Standard Model.}
\label{fig:diffB2Dstar}
\end{center}
\end{figure}
The four panels in Fig.~\ref{fig:diffB2Dstar} illustrate the differential decay distributions of $B\to D^*\ell\nu_{\ell}$​ as functions of 
$q^2$, $\cos\theta$, $\cos\theta_v$ and $\phi$, respectively. The red and blue shaded bands correspond to Standard Model theoretical predictions for 
$\ell=\mu$ and $\ell=\tau$ modes, computed using the combined lattice QCD and LCSR form factor inputs summarised in Table.~\ref{B2DandDstar}, while 
the vertical black error bars denote the binned experimental measurements reported by the Belle~II Collaboration~\cite{Belle-II:2023okj} in 2023. 
Overall, the central values of the theoretical predictions for the muonic channel are broadly consistent with experimental data within quoted 
uncertainties across most of the kinematic space, though mild local deviations can be observed in certain angular regions. If these local tensions 
persist upon improved theoretical and experimental precision, plausible explanations include underestimated systematic uncertainties in the 
lattice QCD and LCSR form factor parametrisations, residual shape biases introduced by the BGL $z$-series interpolation procedure, unaccounted 
higher-order QCD radiative corrections in the theoretical framework, or potential experimental reconstruction efficiency corrections that require 
further validation. At the present precision, however, the observed discrepancies do not provide compelling evidence for physics beyond the 
Standard Model.

\begin{figure}[htbp]
\begin{center}
\includegraphics[width=0.8 \columnwidth]{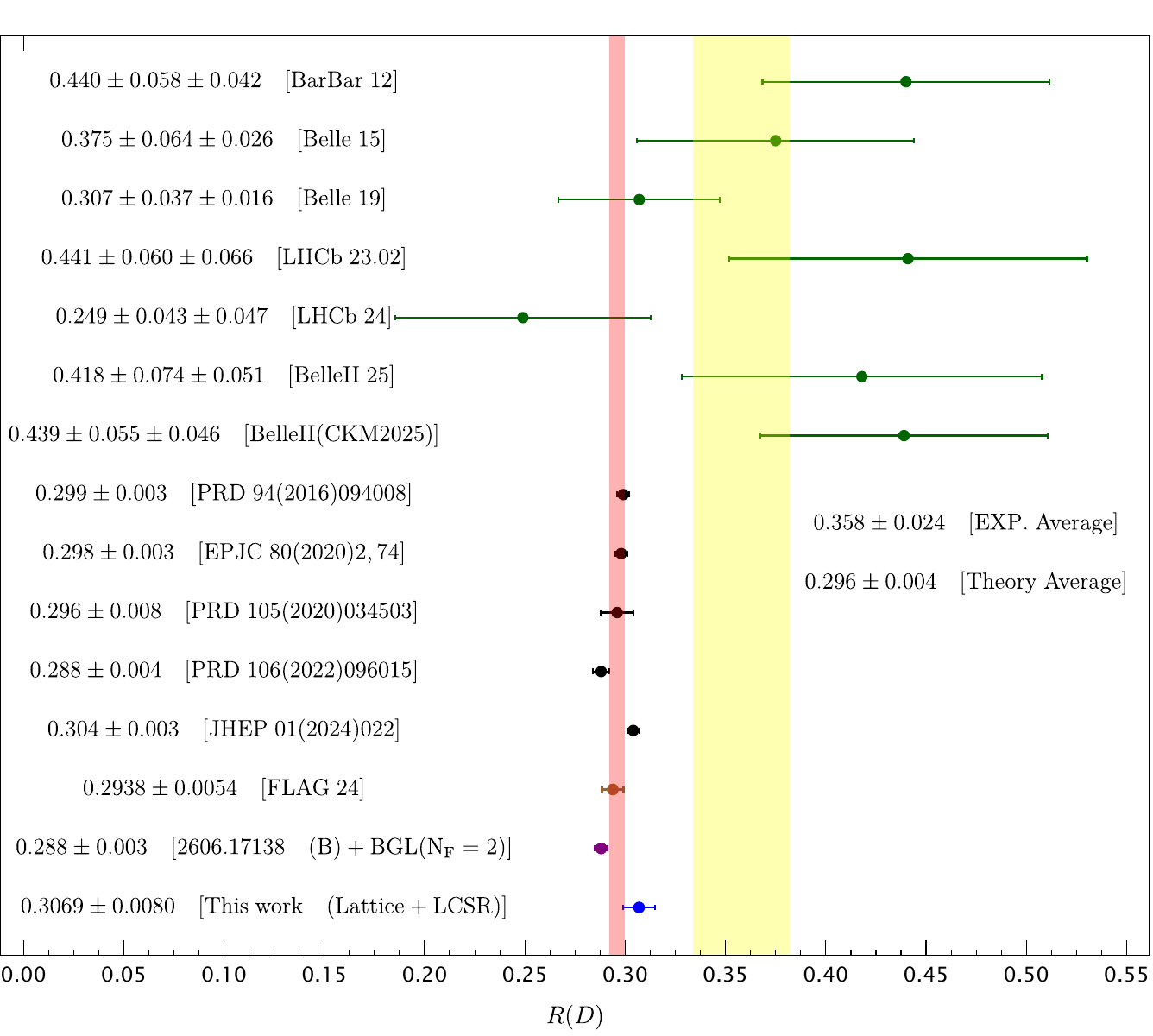}
\hspace{1.0 cm}
\caption{Summary of experimental and theoretical determinations of $R(D)$. Green horizontal error bars correspond to experimental 
measurements~\cite{BaBar:2012obs,BaBar:2013mob,Belle:2015qfa,Belle:2019rba,LHCb:2023zxo,LHCb:2024jll,Belle-II:2025yjp,Tsaklidis:CKM2025}, 
with central values and uncertainties displayed next to each entry; the yellow shaded band marks the experimental average of these results. 
Black error bars denote various theoretical predictions~\cite{Bigi:2016mdz,Bordone:2019vic,Martinelli:2021onb,Bernlochner:2022ywh,Ray:2023xjn}, 
and the red shaded band indicates the arithmetic average of these theoretical results. The brown error bar corresponds to the lattice average from 
FLAG 2024~\cite{FLAG:2024oxs}. The purple error bar shows the prediction from the (B)+BGL($N_f=2$) fit scenario in Ref.~\cite{Fang:2026hru}, 
and the blue error bar represents the result derived from the combined lattice QCD and LCSR fit scenario listed in Table.~\ref{B2DandDstar} 
of the present work.}
\label{fig:RD}
\end{center}
\end{figure}
To intuitively compare experimental results and various theoretical evaluations of $R(D)$, alongside lattice QCD strategies for evaluating lepton-flavour-universality ratio
\begin{equation}
R(D)=\frac{\mathcal{BR}(B\to D\tau\nu_{\tau})}{\mathcal{BR}(B\to D\mu\nu_{\mu})}\;,
\end{equation}
we then summarize all available determinations in Fig.~\ref{fig:RD}. As illustrated in Fig.\ref{fig:RD}, the experimental measurements of $R(D)$ 
from the BaBar, Belle, Belle II and LHCb collaborations~\cite{BaBar:2012obs,BaBar:2013mob,Belle:2015qfa,Belle:2019rba,LHCb:2023zxo,LHCb:2024jll,Belle-II:2025yjp,Tsaklidis:CKM2025} 
(green error bars) exhibit noticeable scatter, and their experimental average, indicated by the yellow shaded band, lies visibly higher than the 
arithmetic average of diverse Standard Model theoretical 
predictions~\cite{Bigi:2016mdz,Bordone:2019vic,Martinelli:2021onb,Bernlochner:2022ywh,Ray:2023xjn} (red shaded band). This well-known tension 
between experimental observations and Standard Model calculations persists and constitutes one of the long standing anomalies in semileptonic 
$B$-meson decays. Multiple theoretical evaluations including LCSR, lattice QCD computations and the FLAG 2024 lattice average produce central values 
clustered around $0.26$-$0.30$, which are systematically lower than the experimental mean. Our result (blue error bar), obtained via the combined 
lattice QCD and SCET sum rule fitting scheme listed in Table.~\ref{B2DandDstar}, yields $R(D)=0.3069\pm0.0080$. Its central value is consistent 
within uncertainties with most existing theoretical predictions, and sits slightly above the majority of pure lattice QCD outcomes such as the 
FLAG 2024~\cite{FLAG:2024oxs} average and the (B)+BGL($N_f​=2$) result. Compared with the broad experimental band, our prediction still deviates 
from the experimental average, maintaining the global theory--experiment tension for $R(D)$. Benefiting from the joint constraint of lattice QCD 
and LCSR inputs, this combined analysis offers a more complete theoretical evaluation of $R(D)$ with reliably quantified uncertainties, providing 
an independent Standard Model prediction for ongoing and future experimental tests of flavour-universality violation. If future precise 
Belle II measurements continue to favor the higher experimental central region, the discrepancy with Standard Model theory would persist and 
strengthen the case for potential new physics contributions in charged current $B\to D\ell\nu_{\ell}$ transitions. On the other hand, improved 
control over lattice systematic uncertainties and refined LCSR inputs will be essential to solidify the theoretical band and clarify whether 
the observed tension originates from underestimated theoretical uncertainties or genuine physics beyond the Standard Model.

\begin{figure}[htbp]
\begin{center}
\includegraphics[width=0.8 \columnwidth]{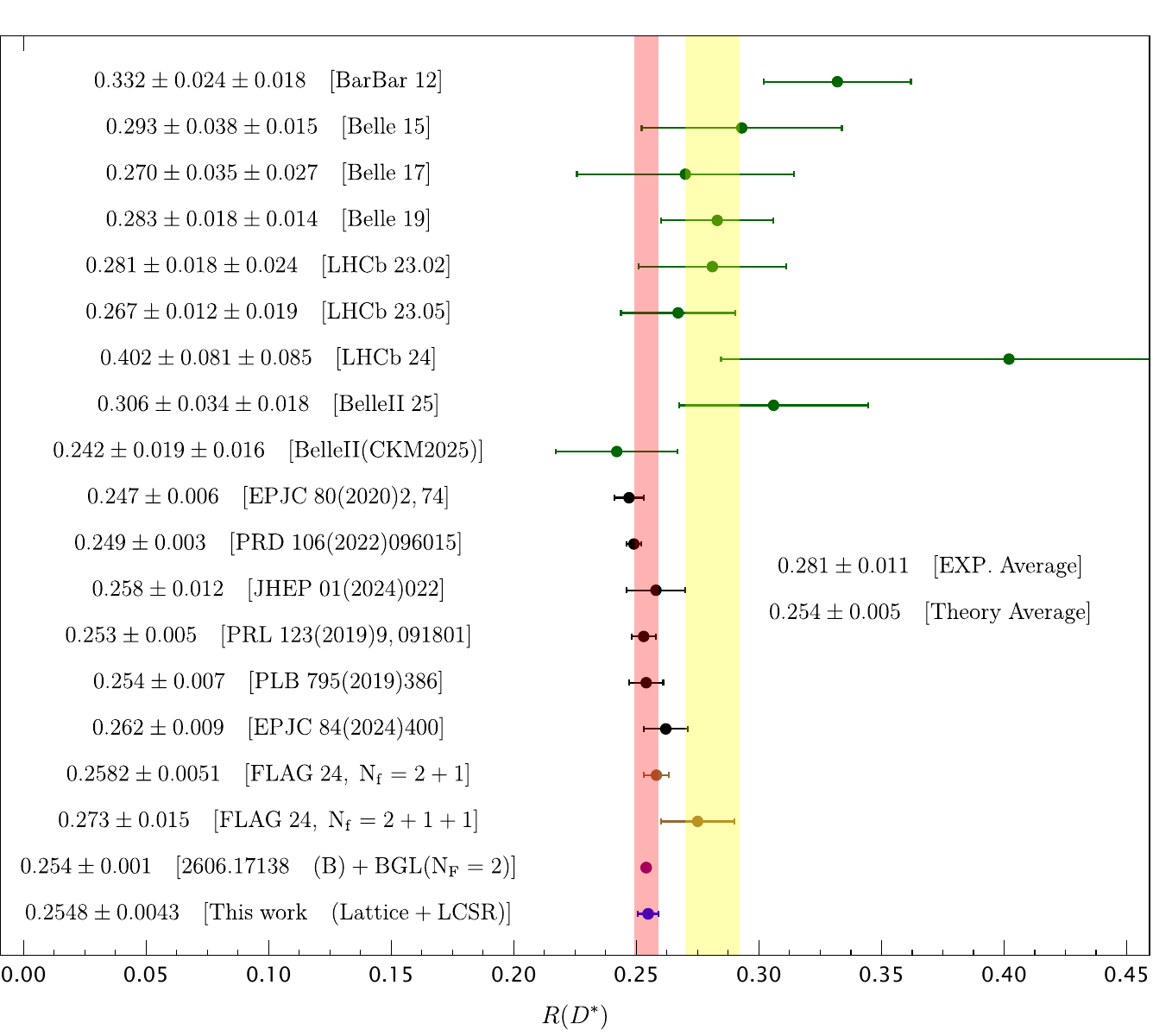}
\hspace{1.0 cm}
\caption{Summary of experimental and theoretical determinations of $R(D^*)$. Green error bars correspond to experimental 
measurements~\cite{BaBar:2012obs,BaBar:2013mob,Belle:2015qfa,Belle:2019rba,LHCb:2023zxo,LHCb:2024jll,Belle-II:2025yjp,Belle:2016dyj,Belle:2017ilt,LHCb:2023uiv,Tsaklidis:CKM2025}, 
with central values and uncertainties displayed next to each entry; the yellow shaded band marks the experimental average of these results. 
Black error bars denote various theoretical 
predictions~\cite{Bordone:2019vic,Bernlochner:2022ywh,Ray:2023xjn,BaBar:2019vpl,Gambino:2019sif,Martinelli:2023fwm}, 
and the red shaded band indicates the arithmetic average of these theoretical results. The brown error bar corresponds to the lattice average from 
FLAG 2024~\cite{FLAG:2024oxs}. The purple error bar shows the prediction from the (B)+BGL($N_f=2$) fit scenario in Ref.~\cite{Fang:2026hru}, 
and the blue error bar represents the result derived from the combined lattice QCD and LCSR fit scenario listed in Table.~\ref{B2DandDstar} 
of the present work.}
\label{fig:RDstar}
\end{center}
\end{figure}
Following the same strategy adopted for $R(D)$, we summarize experimental measurements and theoretical predictions for the lepton-flavour-universality ratio
\begin{equation}
R(D^*)=\frac{\mathcal{BR}(B\to D^*\tau\nu_{\tau})}{\mathcal{BR}(B\to D^*\mu\nu_{\mu})}\;,
\end{equation}
in Fig.~\ref{fig:RDstar} to enable direct comparisons 
among available experimental results, general theoretical evaluations and dedicated lattice QCD computations. The experimental results 
(green error bars) exhibit considerable dispersion, leading to an experimental average of $0.281\pm 0.011$. Theoretical calculations including 
lattice QCD and LCSR approaches (black error bars) are tightly grouped around smaller values, yielding a theoretical average $0.254\pm 0.005$. 
Visible tension persists between the ensemble of experimental measurements and Standard Model theoretical evaluations of $R(D^* )$. Our result, 
$0.2548\pm 0.0043$, obtained from the combined correlated SCET sum rule and lattice QCD fit strategy, is in excellent agreement with the overall 
theoretical average. Numerically, our prediction is consistent with the FLAG 2024~\cite{FLAG:2024oxs} lattice averages and the (B)+BGL(Nf​=2) 
result~\cite{Fang:2026hru} from recent phenomenological analysis, while it lies systematically below the central region of the experimental average. 
Taken together with the finding for $R(D)$, this outcome reinforces the long-standing $R(D^*)$ anomaly, experimental determinations of both 
observables tend toward larger values than Standard Model expectations. Further reduction of theoretical systematic uncertainties and 
higher precision Belle II measurements will be critical to clarify whether the observed tension originates from underestimated theoretical 
uncertainties or signals new physics in charged-current $B\to D^{(*)}\ell\nu_{\ell}$ decays.

\begin{figure}[t]
\begin{center}
\includegraphics[width=0.8 \columnwidth]{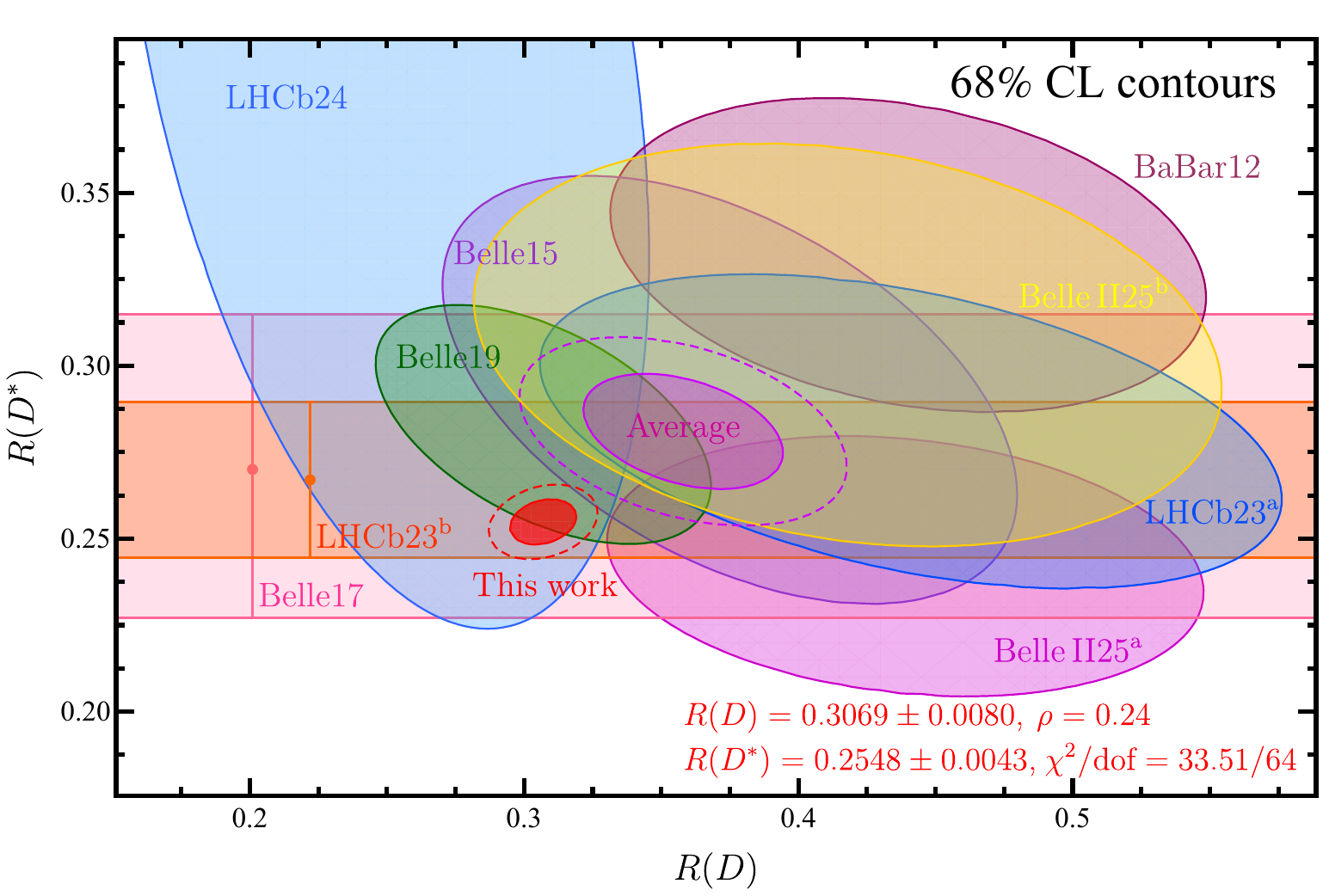}
\hspace{1.0 cm}
\caption{The correlated theoretical predictions for $R(D)$ and $R(D^*)$ obtained from our combined BGL fit incorporating lattice QCD inputs and 
LCSR results. Experimental measurements reported by the BaBar~\cite{BaBar:2012obs,BaBar:2013mob}, 
Belle~\cite{Belle:2015qfa,Belle:2019rba,Belle:2016dyj,Belle:2017ilt}, Belle II~\cite{Belle-II:2025yjp} and
LHCb~\cite{LHCb:2023zxo,LHCb:2024jll,LHCb:2023uiv} collaborations are overlaid for comparison. 
The red solid and dashed ellipses denote the $68\%$ and $95.45\%$ confidence level contours derived in the present work, respectively.}
\label{fig:RD-RDS}
\end{center}
\end{figure}
Complementing the separate analyses of $R(D)$ and $R(D^*)$, Fig.~\ref{fig:RD-RDS} illustrates the two-dimensional correlated constraints on the 
($R(D),R(D^*)$) plane. The colored elliptical contours correspond to the $68\%$ confidence-level experimental measurements from the 
 BaBar~\cite{BaBar:2012obs,BaBar:2013mob}, 
Belle~\cite{Belle:2015qfa,Belle:2019rba,Belle:2016dyj,Belle:2017ilt}, Belle II~\cite{Belle-II:2025yjp} and
LHCb\cite{LHCb:2023zxo,LHCb:2024jll,LHCb:2023uiv} collaborations. One clearly observes substantial scatter among different experimental datasets, 
and the combined experimental average lies systematically above most Standard Model theoretical predictions. Our correlated theoretical result, 
represented by the red solid and dashed ellipses for the $68\%$ and $95.45\%$ CL regions respectively, is derived from a joint BGL fit combining 
lattice QCD and correlated SCET sum rule data points. The central values of our prediction are located at relatively low $R(D)$ and $R(D^*)$ compared 
with the bulk of experimental allowed regions. There exists only modest overlap between our theoretical contour and many experimental ellipses. Meanwhile, our result shows a substantial deviation from the experimental average at $68\%$ CL
which visually demonstrates the persistent tension between Standard Model expectations and experimental observations for semileptonic 
$B\to D^{(*)}\ell\nu_{\ell}$ decays. Thanks to the simultaneous extraction of $R(D)$ and $R(D^*)$ with properly treated correlation between 
these two observables, our work provides a two-dimensional Standard Model prediction. This correlated prediction offers a more stringent test of
lepton-flavour-universality and serves as an important reference for future experimental analyses and potential searches for new physics effects beyond 
the Standard Model.

\section{Conclusions}\label{sec:conclusion}
In this work, we perform global joint fits of the model-independent BGL dispersive parameterization for semileptonic $B\to D^{(*)}\ell\nu_{\ell}$​ 
decays, integrating three complementary categories of state-of-the-art inputs: the latest lattice QCD form factor simulations from FNAL/MILC, HPQCD 
and JLQCD collaborations, correlated SCET sum rule predictions including full NLO QCD and NLP corrections at large hadronic recoil, and binned 
differential decay spectra plus angular observables released by Belle and Belle II. We further impose strong unitarity bounds derived from four 
heavy-to-heavy transition channels with HQET power corrections up to $\mathcal{O}(\alpha_{s},\epsilon_{b},\epsilon_{c},\epsilon_c^2)$, which effectively constrain the space of 
BGL expansion coefficients and suppress uncontrolled systematic uncertainties from $z$-series extrapolation. We carry out three nested 
fit scenarios with progressively enlarged theoretical and experimental constraints: lattice-only fits, combined lattice QCD and LCSR theory fits, 
and full global fits supplemented by Belle(II) experimental data. By comparing the extracted BGL shape parameters across the three scenarios, 
we verify that incorporating correlated large-recoil LCSR predictions dramatically shrinks the uncertainties of high-order form-factor coefficients, 
and the inclusion of precise Belle II decay observables yields further moderate tightening on the parameter space. From the complete global fit 
combining all theoretical and experimental inputs, we extract the CKM matrix element $|V_{cb}​|=(39.18 \pm 0.47)\times 10^{-3}$ via simultaneous 
constraints from both $B\to D\ell\nu_{\ell}$​ and $B\to D^*\ell\nu_{\ell}$ channels. Our determination of $|V_{cb}|$ has a smaller overall uncertainty than many previous exclusive determinations, and the central value differs moderately from previous Bayesian analyses owing to differences in the theoretical and experimental inputs and in the statistical inference frameworks. In particular, Ref.~\cite{Fang:2026hru} employs Bayesian MCMC sampling, while Ref.~\cite{Bordone:2026gqs} uses Bayesian nested sampling implemented with \texttt{PyMultiNest}, whereas the present analysis adopts constrained frequentist $\chi^2$ minimization. Using the fitted form factor coefficients from the lattice QCD and LCSR joint 
theoretical fit, we further derive updated Standard Model predictions for the lepton-flavour-universality ratios $R(D)=0.3069\pm0.0080$ and 
$R(D^*)=0.2548\pm0.0043$, alongside full kinematic distributions of differential decay widths $d\Gamma/dq^2$ and angular observables for both muonic 
and tauonic final states. Summaries of published experimental and theoretical determinations of $R(D)$ and $R(D^*)$ confirm the well-known persistent 
tension, that is the experimental world averages for both observables lie systematically above the aggregate Standard Model theoretical band. 
Our central value for $R(D^*)$ aligns well with the global theory average, while our $R(D)$ prediction sits slightly higher than pure lattice averages 
from FLAG 2024. We further construct the two-dimensional correlated $68\%$ CL contour for $(R(D),R(D^*))$ based on our correlated form factor fit, 
which is presented alongside all experimental elliptical constraints. Our red theoretical contour shows a substantial deviation from the $68\%$ CL 
experimental average with nearly no overlapping region, visually demonstrating the unresolved theory-experiment discrepancy for semileptonic 
$B\to D^{(*)}\ell\nu_{\ell}$​ decays. At present precision, mild local deviations exist between our theoretical angular distributions and Belle II 
measurements in specific kinematic regions, yet no statistically significant evidence for new physics can be established. The observed tension may 
originate from underestimated lattice/LCSR systematic uncertainties, residual shape biases in BGL $z$-expansion, missing higher-order(-power) 
corrections, or unvalidated experimental reconstruction efficiencies. Moving forward, further precision improvements in lattice QCD non-zero-recoil form 
factor calculations, refined next-to-next-to-leading-order or uninvestigated next-to-leading-power corrections, and high-statistics Belle II 
measurements will be critical to disentangling theoretical systematic effects from potential charged-current new physics contributions in 
$b\to c\ell\nu_{\ell}$​ transitions. Our complete set of correlated SM predictions for form factors, $|V_{cb}|$, differential decay spectra, 
angular observables and the two-dimensional $R(D)$ and $R(D^*)$ plane provides a robust, up-to-date reference for subsequent flavour physics investigations.

\subsection*{Acknowledgements}
This work was supported in part by National Natural Science Foundation of China under Contract No.12405109, 12275036;
in part by Natural Science Foundation of Chongqing under Contract No.CSTB2025NSCQ-GPX0919, CSTB2025NSCQ-GPX0945, 
cstc2021jcyj-msxmX0681, cstc2021jcyj-msxmX0678;
in part by the Science and Technology Research Program of Chongqing Municipal Education Commission under Grant No. KJQN202201527;
in part by Research Foundation of Chongqing University of Science and Technology under Grant No.ckrc20231220.
Xue-Chen Zhao is supported by the China Postdoctoral Science Foundation under Grant Number GZB20250794, 2026T190897 and 2026M793702.

\appendix
\section*{Appendix}  
\section{Input parameters}\label{Input parameters}
\begin{table}[htbp]
	\centering
	\renewcommand{\arraystretch}{1.2}
	\begin{tabular}{crrrr}
		\hline\hline
		$F_{j}$ & \multicolumn{1}{c}{$A_{j}$} & \multicolumn{1}{c}{$B_{j}$} & \multicolumn{1}{c}{$C_{j}$} & \multicolumn{1}{c}{$D_{j}$} \\
		\hline
		$S_{1}$ & $0.9813$ & $0.0409$ & $-0.0443$ & $0.0638$ \\
		$S_{2}$ & $0.9460$ & $-0.0769$ & $0.0921$ & $0.0019$ \\
		$S_{3}$ & $0.9460$ & $-0.0407$ & $0.0363$ & $0.0072$ \\
		\hline
		$P_{1}$ & $1.0503$ & $-0.2363$ & $0.1454$ & $-0.0353$ \\
		$P_{2}$ & $0.9045$ & $-0.0764$ & $0.0554$ & $-0.0112$ \\
		$P_{3}$ & $0.9305$ & $-0.1479$ & $0.1283$ & $-0.0229$ \\
		\hline
		$V_{1}$ & \multicolumn{1}{c}{$1$} & \multicolumn{1}{c}{$0$} & \multicolumn{1}{c}{$0$} & \multicolumn{1}{c}{$0$} \\
		$V_{2}$ & $0.9676$ & $-0.1328$ & $0.1284$ & $-0.0209$ \\
		$V_{3}$ & $1.0326$ & $-0.2828$ & $0.1994$ & $-0.0426$ \\
		$V_{4}$ & $1.0640$ & $-0.0914$ & $0.1214$ & $-0.0162$ \\
		$V_{5}$ & $1.0907$ & $-0.0436$ & $0.0972$ & $-0.0091$ \\
		$V_{6}$ & $1.3313$ & $-0.2985$ & $0.1734$ & $-0.0451$ \\
		$V_{7}$ & $1.1878$ & $-0.1421$ & $0.1199$ & $-0.0244$ \\
		\hline
		$A_{1}$ & $0.8925$ & $-0.0083$ & $0.0414$ & $0.0257$ \\
		$A_{2}$ & $0.9278$ & $0.0279$ & $-0.0227$ & $0.0534$ \\
		$A_{3}$ & $0.8925$ & $-0.0045$ & $0.0510$ & $0.0112$ \\
		$A_{4}$ & $0.8925$ & $-0.0280$ & $0.0824$ & $0.0076$ \\
		$A_{5}$ & $0.8925$ & $0.1774$ & $-0.0721$ & $0.0081$ \\
		$A_{6}$ & $0.9278$ & $-0.1282$ & $0.0403$ & $0.0694$ \\
		$A_{7}$ & $0.8925$ & $-0.0567$ & $0.0826$ & $0.0057$ \\
		\hline\hline
	\end{tabular}	
	\caption{Coefficients for the expansion of the ratios $R_{j}(\omega)$ in equation~\eqref{eq_Rj}, considering the $\mathcal{O}(\alpha_{s})$ perturbative corrections as well as $\mathcal{O}(\varepsilon_{c}),\mathcal{O}(\varepsilon_{b}),\mathcal{O}(\varepsilon_{c}^{2})$ power corrections.}
	\label{Rj}
\end{table}

\section{Correlation matrix}\label{Correlation matrix}
\begin{sidewaystable}[p] 
	\centering
	\renewcommand{\arraystretch}{1.2}
	\begin{tabular}{|c|cccccccccccccccc|}
		\hline
~ &  \multicolumn{15}{|c|}{Correlation matrix} \\ \hline
$b_0^{f_+}$ & 1 & 0.22 & 0.13 & 0.13 & 0.17 & 0 & 0 & 0 & 0 & 0 & 0 & 0 & 0 &0 & 0 \\
$b_1^{f_+}$ &  & 1 & -0.45 & 0.83 & -0.40 & 0 & 0 & 0 & 0 & 0 & 0 & 0 & 0 & 0 & 0 \\
$b_2^{f_+}$ &  &  & 1 & -0.49 & 0.99 & 0 & 0 & 0 & 0 & 0 & 0 & 0 & 0 & 0 & 0 \\
$b_1^{f_0}$ &  &  &  & 1 & -0.50 & 0 & 0 & 0 & 0 & 0 & 0 & 0 & 0 & 0 & 0 \\
$b_2^{f_0}$ &  &  &  &  & 1 & 0 & 0 & 0 & 0 & 0 & 0 & 0 & 0 & 0 & 0 \\
$b_0^{g}$ &  &  &  &  &  & 1 & 0.03 & 0.04 & 0.14 & 0.02 & 0.00 & 0.06 & 0.03 & 0.02 & 0.00 \\
$b_1^{g}$ &  &  &  &  &  &  & 1 & -0.72 & -0.03 & 0.16 & 0.00 & 0.08 & 0.01 & 0.16 & -0.04 \\
$b_2^{g}$ &  &  &  &  &  &  &  & 1 & 0.03 & -0.03 & 0.12 & -0.02 & 0.06 & -0.03 & 0.07 \\
$b_0^{f}$ &  &  &  &  &  &  &  &  & 1 & -0.04 & 0.07 & 0.01 & 0.05 & 0.07 & 0.00 \\
$b_1^{f}$ &  &  &  &  &  &  &  &  &  & 1 & -0.48 & 0.61 & -0.31 & 0.53 & -0.33 \\
$b_2^{f}$ &  &  &  &  &  &  & &  & & & 1 & -0.35 & 0.60 & -0.23 & 0.60 \\
$b_1^{F_1}$ &  &  &  &  &  & &  &  &  &  &  & 1 & -0.36 & 0.66 & -0.34 \\
$b_2^{F_1}$ &  &  &  &  &  &  & & &  &  & &  & 1 & -0.10 & 0.92 \\
$b_1^{F_2}$ &   &  &  &  &  &  &  &  &  &  &  &  & & 1 & -0.36 \\
$b_2^{F_2}$ &   &  &  &  &  &  &  &  &  &  &  &  &  &  & 1 \\
		\hline
	\end{tabular}	
	\caption{Correlation matrix of the fitted parameters obtained under the lattice‑only scenario.}
	\label{CorrelationLattice}
\end{sidewaystable}

\begin{sidewaystable}[p] 
	\centering
	\renewcommand{\arraystretch}{1.2}
	\begin{tabular}{|c|cccccccccccccccc|}
		\hline
~ &  \multicolumn{15}{|c|}{Correlation matrix} \\ 
\hline
$b_0^{f_+}$ & 1& 0.27 & -0.06 & 0.20 & 0.02 & 0.01 & 0.01 & 0.02 & 0.02 & 0.01 & 0.03 & 0.02 & 0.03 & 0.03 & 0.02 \\
$b_1^{f_+}$ &  & 1. & -0.47 & 0.82 & -0.38 & 0.00 & 0.00 & -0.01 & 0.00 & 0.00 & -0.01 & 0.01 & -0.01 & 0.01 & -0.01 \\
$b_2^{f_+}$ &  &  & 1 & -0.44 & 0.98 & 0.05 & 0.03 & 0.08 & 0.04 & 0.03 & 0.13 & 0.05 & 0.14 & 0.06 & 0.10 \\
$b_1^{f_0}$ & &  & & 1 & -0.47 & -0.01 & 0.00 & -0.02 & 0.00 & 0.00 & -0.02 & 0.00 & -0.02 & 0.00 & -0.02 \\
$b_2^{f_0}$ &   & & &  & 1 & 0.05 & 0.03 & 0.08 & 0.05 & 0.03 & 0.13 & 0.05 & 0.14 & 0.07 & 0.10 \\
$b_0^{g}$ & &  & & &  & 1 & 0.09 & -0.13 & 0.14 & 0.00 & 0.11 & 0.03 & 0.07 & -0.01 & 0.03 \\
$b_1^{g}$ &   &  &  &  &  &  & 1 & -0.65 & -0.03 & 0.23 & -0.02 & 0.12 & 0.00 & 0.19 & -0.11 \\
$b_2^{g}$ &  &  &  &  &  &  &  & 1 & 0.04 & -0.12 & 0.32 & -0.09 & 0.16 & -0.1 & 0.22 \\
$b_0^{f}$ &  &  &  &  & & & &  & 1& -0.04 & 0.03 & 0.01 & 0.01 & 0.06 & -0.06 \\
$b_1^{f}$ &  &  &  &  &  &  & &  &  & 1 & -0.5 & 0.59 & -0.36 & 0.51 & -0.37 \\
$b_2^{f}$ &   &  &  &  &  &  &  &  &  &  & 1 & -0.34 & 0.6 & -0.22 & 0.53 \\
$b_1^{F_1}$ &  &  &  &  &  &  & &  &  &  &  & 1 & -0.53 & 0.64 & -0.45 \\
$b_2^{F_1}$ &  &  &  &  &  &  &  &  &  &  &  &  & 1 & -0.19 & 0.77 \\
$b_1^{F_2}$ &   &  &  &  &  &  &  &  &  &  &  &  &  & 1 & -0.61 \\
$b_2^{F_2}$ &  &  & & & & & & & & & & & &  & 1 \\
		\hline
	\end{tabular}	
	\caption{Correlation matrix of the fitted parameters obtained under the lattice$\oplus$LCSR scenario.}
	\label{CorrelationLatticeLCSR}
\end{sidewaystable}

\begin{sidewaystable}[p] 
	\centering
	\renewcommand{\arraystretch}{1.2}
	\begin{tabular}{|c|ccccccccccccccccc|}
		\hline
~ &  \multicolumn{16}{|c|}{Correlation matrix} \\ 
\hline
$|V_{cb}|$ & 1 & -0.28 & -0.18 & -0.14 & -0.13 & -0.19 & -0.19 & -0.07 & -0.01 & -0.55 & -0.28 & 0.03 & -0.27 & 0.13 & -0.20 & 0.15 \\
$b_0^{f_+}$ & ~ & 1 & 0.20 & -0.26 & 0.15 & -0.16 & 0.06 & 0.02 & 0.00 & 0.16 & 0.08 & -0.01 & 0.08 & -0.04 & 0.06 & -0.05 \\
$b_1^{f_+}$ & ~ & ~ & 1 & -0.72 & 0.81 & -0.61 & 0.04 & 0.02 & -0.01 & 0.10 & 0.05 & -0.01 & 0.05 & -0.03 & 0.04 & -0.03 \\
$b_2^{f_+}$ & ~ & ~ & ~ & 1 & -0.61 & 0.93 & 0.02 & 0.00 & 0.02 & 0.08 & 0.03 & 0.01 & 0.03 & -0.01 & 0.02 & -0.01 \\
$b_1^{f_0}$ & ~ & ~ & ~ & ~ & 1 & -0.69 & 0.03 & 0.01 & -0.01 & 0.07 & 0.04 & -0.01 & 0.04 & -0.02 & 0.03 & -0.02 \\
$b_2^{f_0}$ & ~ & ~ & ~ & ~ & ~ & 1 & 0.03 & 0.01 & 0.02 & 0.11 & 0.05 & 0.00 & 0.05 & -0.02 & 0.03 & -0.02 \\
$b_0^{g}$ & ~ & ~ & ~ & ~ & ~ & ~ & 1 & -0.24 & -0.16 & 0.26 & 0.00 & -0.06 & 0.10 & -0.05 & -0.01 & -0.04 \\
$b_1^{g}$ & ~ & ~ & ~ & ~ & ~ & ~ & ~ & 1 & -0.76 & -0.01 & 0.19 & -0.19 & 0.11 & -0.06 & 0.20 & -0.17 \\
$b_2^{g}$ & ~ & ~ & ~ & ~ & ~ & ~ & ~ & ~ & 1 & 0.04 & -0.15 & 0.17 & -0.05 & -0.02 & -0.13 & 0.11 \\
$b_0^{f}$ &  ~ & ~ & ~ & ~ & ~ & ~ & ~ & ~ & ~ & 1 & -0.12 & 0.07 & -0.10 & 0.04 & 0.00 & -0.06 \\
$b_1^{f}$ &  ~ & ~ & ~ & ~ & ~ & ~ & ~ & ~ & ~ & ~ & 1 & -0.81 & 0.55 & -0.41 & 0.43 & -0.34 \\
$b_2^{f}$ &  ~ & ~ & ~ & ~ & ~ & ~ & ~ & ~ & ~ & ~ & ~ & 1 & -0.39 & 0.30 & -0.34 & 0.29 \\
$b_1^{F_1}$ & ~ & ~ & ~ & ~ & ~ & ~ & ~ & ~ & ~ & ~ & ~ & ~ & 1 & -0.93 & 0.44 & -0.54 \\
$b_2^{F_1}$ & ~ & ~ & ~ & ~ & ~ & ~ & ~ & ~ & ~ & ~ & ~ & ~ & ~ & 1 & -0.37 & 0.55 \\
$b_1^{F_2}$ & ~ & ~ & ~ & ~ & ~ & ~ & ~ & ~ & ~ & ~ & ~ & ~ & ~ & ~ & 1 & -0.89 \\
$b_2^{F_2}$ & ~ & ~ & ~ & ~ & ~ & ~ & ~ & ~ & ~ & ~ & ~ & ~ & ~ & ~ & ~ & 1 \\
		\hline
	\end{tabular}	
	\caption{Correlation matrix of the fitted parameters obtained under the lattice$\oplus$LCSR$\oplus$Exp. scenario.}
	\label{CorrelationALL}
\end{sidewaystable}

\FloatBarrier

\bibliographystyle{apsrev4-1}
\bibliography{references}
\end{document}